\documentclass[11pt]{article}

\usepackage{hyperref}
\usepackage{graphicx}
\usepackage{amsmath,amssymb,amsfonts,dsfont,slashed}
\usepackage{cite}
\usepackage{booktabs}

\allowdisplaybreaks[1]

\newcommand{\beq}{\begin{equation}}
\newcommand{\eeq}{\end{equation}}
\newcommand{\beqa}{\begin{eqnarray}}
\newcommand{\eeqa}{\end{eqnarray}}
\newcommand{\bal}{\begin{align}}
\newcommand{\Fpi}{F_\pi}
\newcommand{\mpi}{M_{\pi}}
\newcommand{\mpii}{M_{\pi^0}}

\newcommand{\diff}{\text{d}}
\newcommand{\eps}{\epsilon}

\newcommand{\F}{\mathcal{F}}
\newcommand{\Order}{\mathcal{O}}

\renewcommand{\Im}{\text{Im}\,}

\newcommand{\Imt}{\text{Im}_t\,}
\newcommand{\Ims}{\text{Im}_s\,}
\newcommand{\Imspipi}{\text{Im}_s^{\pi\pi}\,}

\newcommand{\Tr}{\text{Tr}}
\newcommand{\Sym}{\mathcal{S}}
\newcommand{\lam}{\lambda_{12}}

\def\XXint#1#2#3{{\setbox0=\hbox{$#1{#2#3}{\int}$}
     \vcenter{\hbox{$#2#3$}}\kern-0.5\wd0}}

\begin{document}

\setlength{\unitlength}{1mm}

\numberwithin{equation}{section}

\author{G.~Colangelo, M.~Hoferichter, M.~Procura, and P.~Stoffer}

\title{\bf Dispersive approach to hadronic light-by-light scattering}

\date{}

\maketitle

\begin{center}
{\small
{\it Albert Einstein Center for Fundamental Physics,
Institute for Theoretical Physics, \\
University of Bern, Sidlerstrasse 5, CH--3012 Bern, Switzerland}
}
\end{center}

\bigskip

\begin{abstract}
  Based on dispersion theory, we present a formalism for a
  model-independent evaluation of the hadronic light-by-light contribution
  to the anomalous magnetic moment of the muon. In particular, we comment
  on the definition of the pion pole in this framework and provide a master
  formula that relates the effect from $\pi\pi$ intermediate states to the
  partial waves for the process $\gamma^*\gamma^*\to\pi\pi$. All
  contributions are expressed in terms of on-shell form factors and
  scattering amplitudes, and as such amenable to an experimental
  determination.
\end{abstract}

\section{Introduction}
\label{sec:introduction}

Hadronic contributions dominate the uncertainty in the Standard-Model
prediction of the anomalous magnetic moment of the muon
$a_\mu=(g-2)_\mu/2$, see e.g.~\cite{JN,Prades:2009tw}.  In view of the next
round of $(g-2)_\mu$ experiments at FNAL and J-PARC aimed at reducing the
experimental error by a factor of $4$, control over these hadronic effects
has to be improved substantially to make sure that experiment and
Standard-Model prediction continue to compete at the same accuracy.

The leading hadronic contribution, hadronic vacuum polarization (HVP), can
be related directly to the total hadronic cross section in $e^+e^-$
scattering and, given a dedicated $e^+e^-$ program, it is expected to allow
for the required improvement, see e.g.~\cite{g-2wp}. 
In contrast, the subleading\footnote{At next-to-leading order also two-loop diagrams with HVP insertions appear~\cite{Calmet:1976kd}. Next-to-next-to-leading-order hadronic contributions have been recently considered in~\cite{Kurz:2014wya,Colangelo:2014qya}.} 
hadronic light-by-light (HLbL) scattering
contribution has so far been evaluated within hadronic models and
frameworks that partially incorporate rigorous constraints from
QCD~\cite{deRafael:1993za,Bijnens:1995cc,BPP95,Bijnens:2001cq,Hayakawa:1995ps,Hayakawa:1996ki,Hayakawa:1997rq,Knecht:2001qg,KN,RamseyMusolf:2002cy,MV,Goecke:2010if}.
In this context, a reliable estimate of the uncertainty associated with HLbL scattering
as well as future reductions thereof appear difficult.
  As an alternative, model-independent
approach to the problem, lattice QCD calculations have been
proposed~\cite{Hayakawa:2005eq}, but it is yet premature to make
predictions about when such calculations will become competitive (for the
present status see~\cite{Blum,g-2wp}).  Accordingly, HLbL scattering will
soon dominate the theory error and thus become the roadblock in fully
exploiting the new $(g-2)_\mu$ measurements.

Here, we propose to use dispersion theory to analyze HLbL scattering,
similarly to what is done for HVP. In this framework, the amplitude is
characterized by its analytic structure, i.e.\ poles and cuts, so that the
relevant quantities are residues and imaginary parts, and thus, by
definition, on-shell form factors and scattering amplitudes. In this way, a
direct correspondence to experimentally accessible quantities can
ultimately be established.  While the advantages are evident, such an
approach has been long sought after.  However, due to the more complicated
structure of HLbL scattering, it has not been possible so far to write down
a formula strictly analogous to that for HVP that includes all possible
hadronic intermediate states. In the present paper we make the first step
in that direction, based on the assumption that the most important
contributions are due to the single- and double-pion intermediate states.
While the former has been analyzed in several papers in a (to a large
extent) model-independent way, this is not the case for the latter.  The
main novelty of this paper is a master formula that explicitly relates the
contribution of two-pion intermediate states to $a_\mu$ to the partial
waves for $\gamma^*\gamma^*\to\pi\pi$.  In particular, within our framework
the issues raised in~\cite{Engel:2012xb,Bijnens:2012an,Engel:2013kda}
concerning the dressing of the pion loop can be settled with input from
experiment.

The outline of the paper is as follows: in Sect.~\ref{sec:framework} we
introduce our dispersive approach, we illustrate it using the example of
the pion pole, commenting on its definition within this picture, and then
collect the necessary notation concerning $\gamma^*\gamma^*\to\pi\pi$
partial waves which will later be needed for the analysis of $\pi\pi$
intermediate states. In Sect.~\ref{sec:master}, we derive a set of
dispersion relations for HLbL scattering, leading to an expression for the
$\pi\pi$ contribution to $a_\mu$ in terms of $\gamma^*\gamma^*\to\pi\pi$
partial waves.  Finally, we offer our conclusions and an outlook in
Sect.~\ref{sec:outlook}. Various details of the calculation are discussed
in the appendices.

\section{Dispersive framework for hadronic light-by-light scattering}
\label{sec:framework}

\subsection{Notation}
\label{sec:notation}

We define the HLbL tensor $\Pi^{\mu\nu\lambda\sigma}$ as 
\beq
\Pi^{\mu \nu \lambda \sigma}\big(q_1,q_2,q_3\big)=i^3 \int d^4x \int d^4y \int d^4z\,
e^{-i(x\cdot q_1+y\cdot q_2 +z\cdot q_3)} \langle 0 | T \big\{ j^\mu(x) j^\nu(y) j^\lambda(z)
j^\sigma(0) \big\} | 0 \rangle,
\eeq
where $j^\mu(x)=\sum_i Q_i \bar q_i(x) \gamma^\mu q_i(x)$, $i=u,d,s$, is
the electromagnetic current ($Q_i$ being the charge of the quark in proton
charge units) and the matrix element is to be evaluated in pure QCD 
(i.e.\ for $\alpha=e^2/(4\pi)=0$). In the calculation of $a_\mu$ we take 
the external photon to couple with the fourth current, and denote 
its momentum by $k = q_1+q_2+q_3$. In addition, we need the above tensor in
the kinematic configuration $k^2=0$. Contracted with the appropriate
polarization vectors this gives the matrix element of the leading-order (in
$\alpha$) hadronic contribution to the reaction
\begin{align}
\label{eq:H1234}
H_{\lambda_1 \lambda_2,\lambda_3 \lambda_4}(s,t,u)
&\equiv
\mathcal{M}(\gamma^*(q_1,\lambda_1)\gamma^*(q_2,\lambda_2)\to\gamma^*(-q_3,\lambda_3)\gamma(k,\lambda_4))\notag \\
&=\eps_\mu(\lambda_1,q_1) \eps_\nu(\lambda_2,q_2)
\eps^*_\lambda(\lambda_3,-q_3) \eps^*_\sigma(\lambda_4, k) \Pi^{\mu \nu
  \lambda \sigma}(q_1,q_2,q_3),  
\end{align}
with Mandelstam variables 
\beq
s=(q_1+q_2)^2=(k-q_3)^2,\qquad
t=(q_1+q_3)^2=(k-q_2)^2,\qquad
u=(q_2+q_3)^2=(k-q_1)^2,
\eeq
and $s$-channel scattering angle
 \beq
 \label{schannel_angle}
 z_s=\cos\theta_s=\frac{s}{\big(s-q_3^2\big)\sqrt{\lambda_{12}}}\bigg(t-u+\frac{\big(q_1^2-q_2^2\big)q_3^2}{s}\bigg),\qquad
  \lambda_{12}=\lambda\big(s,q_1^2,q_2^2\big),
 \eeq
with $\lambda(x,y,z)=x^2+y^2+z^2-2(xy+xz+yz)$ the K\"all\'en function.
For the contribution to $a_\mu$ we only need the derivative with respect to the external photon momentum $k^\sigma$, since by virtue of gauge invariance~\cite{Aldins}
\beq
\Pi_{\mu\nu\lambda\sigma}\big(q_1,q_2,k-q_1-q_2\big)=-k^\rho\frac{\partial}{\partial k^\sigma}\Pi_{\mu\nu\lambda\rho}\big(q_1,q_2,k-q_1-q_2\big).
\eeq
The contribution to $a_\mu$ follows from
\begin{align}
\label{amu_proj}
a_\mu&=\lim_{k\to 0}\Tr\Big\{\big(\slashed{p}+m\big)\Lambda^\rho\big(p',p\big)\big(\slashed{p}'+m\big)\Gamma_\rho\big(p',p\big)\Big\},\notag\\  
 \Gamma_\rho\big(p',p\big)&=e^6\int\frac{\diff^4q_1}{(2\pi)^4}\int\frac{\diff^4q_2}{(2\pi)^4}\frac{1}{q_1^2q_2^2q_3^2}
 \frac{\gamma^\mu\big(\slashed{p}'+\slashed{q_1}+m\big)\gamma^\lambda\big(\slashed{p}-\slashed{q_2}+m)\gamma^\nu}{\big((p'+q_1)^2-m^2\big)
 \big((p-q_2)^2-m^2\big)} k^\sigma\frac{\partial}{\partial k^\rho}\Pi_{\mu\nu\lambda\sigma},
\end{align}
with the projector~\cite{Brodsky:1966mv}
\beq
\Lambda^\rho\big(p',p\big)=\frac{m^2}{k^2\big(4m^2-k^2\big)}\Bigg\{\gamma^\rho+\frac{k^2+2m^2}{m\big(k^2-4m^2\big)}\big(p+p'\big)^\rho\Bigg\}.
\eeq
$m$ denotes the mass of the muon, $p$ and $p'=p-k$ the momenta of the incoming and outgoing muon, respectively,
and we have assumed that $\Pi_{\mu\nu\lambda\sigma}$ is already manifestly gauge invariant and crossing symmetric.
The general relation~\eqref{amu_proj} can be further simplified
using the identity
\beq
\big(\slashed{p}+m\big)\gamma^\rho \big(\slashed{p}'+m\big)
=\big(\slashed{p}+m\big)\bigg[\frac{1}{2m}\big(p+p'\big)^\rho+\frac{i}{2m}\sigma^{\rho\tau}k_\tau\bigg] \big(\slashed{p}'+m\big).
\eeq
Explicit expressions will be given in~\eqref{lblmaster}
and~\eqref{lbl_master_average}.

\subsection{Layout of the dispersive approach}
\label{sec:disprel}

In a dispersive approach one exploits the analytic properties of the matrix
element of interest and reconstructs it completely from information on its
analytic singularities: residues of poles, values along cuts, and
subtraction constants (representing singularities at infinity). Depending
on the complexity of the singularity structure of a given amplitude such a
program can be carried out until the very end (as in the case of form
factors), or lead to integral equations amenable to numerical treatment. In
the worst case the singularity structure may be too complex to allow for
an exact treatment. The HLbL amplitude clearly belongs to the latter class,
unfortunately: it has single poles, cuts in all channels (and
simultaneously in different channels), and in all photon momenta squared, 
as well as anomalous thresholds~\cite{Mandelstam,LMS,HCPS}.

\begin{figure}
\centering
$F_\pi^V\big(q_1^2\big)F_\pi^V\big(q_2^2\big)F_\pi^V\big(q_3^2\big)\times\left[
\hspace{0.2cm}
\raisebox{-1.65cm}{
\includegraphics[width=0.2\linewidth]{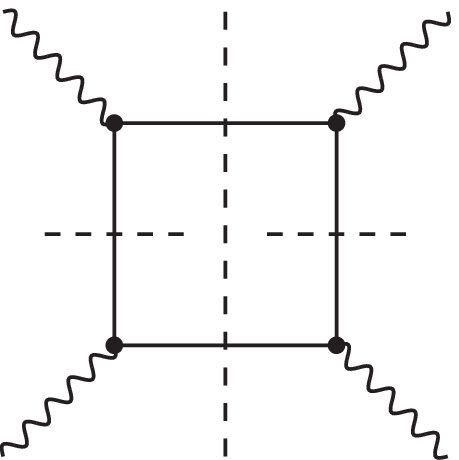} \quad
\includegraphics[width=0.2\linewidth]{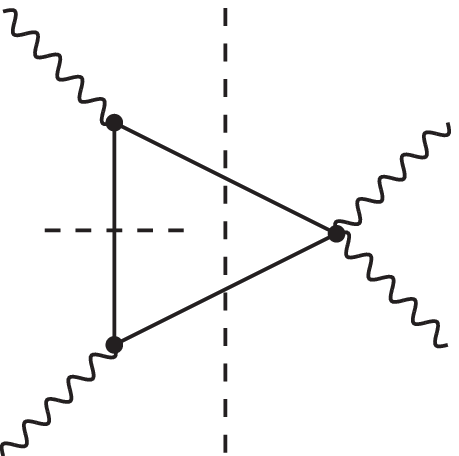}  \quad
\raisebox{0.77cm}{\includegraphics[width=0.2\linewidth]{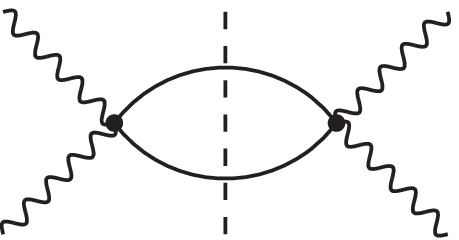}}
}
\hspace{0.2cm}
\right]
$
\caption{Scalar QED diagrams with photon--pion vertices dressed by the
  (appropriate power of) pion vector form factors, in the following
  referred to as FsQED. Solid lines denote pions, wiggly lines photons, and
  the dashed lines indicate the cutting of the pion propagators.}
\label{fig:sQED}
\end{figure}
On the basis of model calculations (see, e.g.~\cite{BPP95}) of the
HLbL contributions to $a_\mu$, it is clear that singularities having higher
thresholds (like the cut due to $\bar K K$ intermediate states) are less
important. It appears therefore reasonable to reduce the complexity of the
problem by limiting ourselves to the lowest-lying intermediate states,
pions,\footnote{We are well aware of the fact that the single poles due to $\eta$, $\eta'$,
  and other higher-mass states are not negligible. They are, however,
  easily taken into account and can be just added to the contributions
  considered here. For the sake of clarity, we limit the discussion to
pions only.} and to allow for at most two pions in intermediate
states. In this approximation the HLbL tensor can be broken down into the
following contributions
\beq
\label{eq:Pibreakdown}
\Pi_{\mu\nu\lambda\sigma}=\Pi_{\mu\nu\lambda\sigma}^{\pi^0\text{-pole}}
+\Pi_{\mu\nu\lambda\sigma}^{\text{FsQED}}+\bar \Pi_{\mu\nu\lambda\sigma}+\cdots,
\eeq
where $\Pi_{\mu\nu\lambda\sigma}^{\pi^0\text{-pole}}$ refers to the pion
pole, $\Pi_{\mu\nu\lambda\sigma}^{\text{FsQED}}$ to the amplitude in scalar QED
with vertices dressed by the (appropriate power of) pion vector form
factors $F_\pi^V(q^2)$ (FsQED), $\bar \Pi_{\mu\nu\lambda\sigma}$ to the remaining
$\pi\pi$ contribution, and the ellipsis 
to higher-mass poles and intermediate states.

The reason for separating the FsQED
contribution from the rest and its precise meaning can be explained as
follows: $\Pi_{\mu\nu\lambda\sigma}^{\text{FsQED}}$ includes the
contribution due to simultaneous two-pion cuts in two of the channels (by
crossing symmetry it contains three contributions with simultaneous
singularities in the $(s,t)$, $(s,u)$, and $(t,u)$ channels, respectively).
One first takes the two-pion cut in the $s$-channel, which gives the
discontinuity as the product of two $\gamma^* \gamma^* \to \pi \pi$
amplitudes, and then selects the Born term (the pure pole term) in each of
the two amplitudes, as illustrated by the leftmost diagram in
Fig.~\ref{fig:sQED}. The singularity of this diagram is therefore given by
four $\pi^+\pi^-\gamma^*$ vertices with on-shell pions---which implies that
these vertices are nothing but the full pion vector form factors. On the
other hand, the singularity structure of this contribution is identical to
that of a Feynman box diagram with four pion propagators: since the four
vertices depend only on the momentum squared of the external photons and on
none of the internal momenta, this contribution is given by the box-diagram
multiplied by three pion vector form factors (since one of the photons is
on-shell). In sQED the box diagram in Fig.~\ref{fig:sQED} is not gauge
invariant on its own, however. The photon--scalar--scalar vertex comes
together with the seagull term (two-photon--two-scalar vertex), with
couplings strictly related to each other: in any amplitude with two or more
photons both vertices have to be taken into account to form a subset of
gauge-invariant diagrams. Therefore, in sQED the box diagram has to be
accompanied by a triangle and a bulb diagram in order to respect gauge
invariance, as shown in Fig.~\ref{fig:sQED}. We do the same here and define
our gauge-invariant box diagram as the charged pion loop calculated within
sQED multiplied by the pion vector form factors.

We stress that the separation of this contribution from the rest is
unambiguous as it is based on its analytic properties, namely the
presence of simultaneous cuts in two channels. The request to have the two
simultaneous cuts is equivalent to putting all pions in the box diagram
on-shell, but does not put constraints on the vertex with the photon, which
is allowed to have its full $q^2$-dependence. The fact that the two
pions in the vertex are both on-shell, however, allows us to identify that
vertex with the pion vector form factor: multiplying the sQED contribution
by three pion form factors, as shown in Fig.~\ref{fig:sQED}, is not an
approximation, but the exact and unambiguous representation of the
contribution with these analytic properties.
How to technically separate this contribution from the others with two-pion
intermediate states and how they contribute to $a_\mu$ will be discussed in
more detail in Sect.~\ref{subsec:master}.

Since we only explicitly consider cuts from up to two-pion intermediate
states, this implies that the analytic structure of the remainder $\bar
\Pi_{\mu\nu\lambda\sigma}$ in~\eqref{eq:Pibreakdown} does not involve
so-called double-spectral regions, i.e.\ parts of the Mandelstam plane with
simultaneous singularities in two Mandelstam variables, and can therefore
be expanded in partial waves, making a dispersive treatment of this part
feasible.  In the rest of this section we first specify the contribution of
the pion pole to $a_\mu$, and then set up the notation for the $\gamma^*
\gamma^* \to \pi \pi$ reaction. Based on these conventions, we will derive
dispersion relations for $\bar \Pi_{\mu\nu\lambda\sigma}$ in
Sect.~\ref{sec:master}.

\subsection{Pion pole}
\label{sec:pion_pole}

The dominant contribution to HLbL scattering at low energy is given by the
$\pi^0$-poles. Their residues are determined by the on-shell, doubly-virtual pion
transition form factor  
$\F_{\pi^0\gamma^*\gamma^*}(q_1^2,q_2^2)$,
which is defined as the current matrix element
\beq
i\int \diff^4 x\, e^{iq\cdot x}\big\langle0\big|T\big\{j_\mu(x)j_\nu(0)\big\}\big|\pi^0(p)\big\rangle=\eps_{\mu\nu\alpha\beta}q^\alpha p^\beta
\F_{\pi^0\gamma^*\gamma^*}\big(q^2,(p-q)^2\big).
\eeq
In these conventions, the $\pi^0$-pole HLbL amplitude reads
\begin{align}
\label{lbl_pi0pole}
 \Pi_{\mu\nu\lambda\sigma}^{\pi^0\text{-pole}}&=\frac{\F_{\pi^0\gamma^*\gamma^*}\big(q_1^2,q_2^2\big)\F_{\pi^0\gamma^*\gamma^*}\big(q_3^2,0\big)}{s-\mpii^2}
 \epsilon_{\mu\nu\alpha\beta}q_1^\alpha q_2^\beta\epsilon_{\lambda\sigma\gamma\delta}q_3^\gamma k^\delta\notag\\
 &+\frac{\F_{\pi^0\gamma^*\gamma^*}\big(q_1^2,q_3^2\big)\F_{\pi^0\gamma^*\gamma^*}\big(q_2^2,0\big)}{t-\mpii^2}
 \epsilon_{\mu\lambda\alpha\beta}q_1^\alpha q_3^\beta\epsilon_{\nu\sigma\gamma\delta}q_2^\gamma k^\delta\notag\\
 &+\frac{\F_{\pi^0\gamma^*\gamma^*}\big(q_2^2,q_3^2\big)\F_{\pi^0\gamma^*\gamma^*}\big(q_1^2,0\big)}{u-\mpii^2}
 \epsilon_{\nu\lambda\alpha\beta}q_2^\alpha q_3^\beta\epsilon_{\mu\sigma\gamma\delta}q_1^\gamma k^\delta.
\end{align}
Its contribution to $a_\mu$ can be derived from
\begin{align}
\label{lblmaster}
 a_\mu&=\frac{1}{48m}\Tr\Big\{\big(\slashed{p}+m\big)\big[\gamma^\rho,\gamma^\sigma\big]\big(\slashed{p}+m\big)\Gamma_{\rho\sigma}\Big\},\notag\\
 \Gamma_{\rho\sigma}&=-e^6\int\frac{\diff^4q_1}{(2\pi)^4}\int\frac{\diff^4q_2}{(2\pi)^4}\frac{1}{q_1^2q_2^2s}
 \frac{\gamma^\mu\big(\slashed{p}+\slashed{q_1}+m\big)\gamma^\lambda\big(\slashed{p}-\slashed{q_2}+m)\gamma^\nu}{\big((p+q_1)^2-m^2\big)
 \big((p-q_2)^2-m^2\big)}\bigg[\frac{\partial}{\partial k^\rho}\Pi_{\mu\nu\lambda\sigma}\bigg]_{k=0}.
\end{align}
This formula holds true if the derivative has a well-defined limit for
$k\to 0$, a condition fulfilled by the $\pi^0$-pole
amplitude~\eqref{lbl_pi0pole}. It follows from~\eqref{amu_proj} by
averaging over the spatial directions of $k$,
see~\cite{Barbieri:1974nc,Barbieri:1974dg,Jegerlehner}, whereupon the limit
may be pulled inside the integral. The final result can be expressed as~\cite{KN}
\begin{align}
\label{amu_pole}
 a_\mu^{\pi^0\text{-pole}}&=-e^6\int\frac{\diff^4q_1}{(2\pi)^4}\int\frac{\diff^4q_2}{(2\pi)^4}
 \frac{1}{q_1^2q_2^2s\big((p+q_1)^2-m^2\big)\big((p-q_2)^2-m^2\big)}\\
 &\times\Bigg\{\frac{\F_{\pi^0\gamma^*\gamma^*}\big(q_1^2,q_2^2\big)\F_{\pi^0\gamma^*\gamma^*}\big(s,0\big)}{s-\mpii^2}T_1(q_1,q_2;p)
 +\frac{\F_{\pi^0\gamma^*\gamma^*}\big(s,q_2^2\big)\F_{\pi^0\gamma^*\gamma^*}\big(q_1^2,0\big)}{q_1^2-\mpii^2}T_2(q_1,q_2;p)\Bigg\},\notag
\end{align}
with
\begin{align}
 T_1(q_1,q_2;p)&=\frac{8}{3} \bigg\{2 p\cdot q_1\,p\cdot q_2\, q_1\cdot q_2+p\cdot q_1\,q_2^2\Big(q_1\cdot q_2+q_1^2-2 p\cdot q_1\Big)
 -\frac{m^2\lambda_{12}}{4}\bigg\},\\
 T_2(q_1,q_2;p)&=\frac{16}{3} \bigg\{p\cdot q_1\Big(p\cdot q_2\, q_1\cdot q_2
 -p\cdot q_1\,q_2^2+(q_1\cdot q_2)^2\Big)-\frac{q_1^2}{2} \Big(3p\cdot q_1\,q_2^2 -p\cdot q_2\, q_1\cdot q_2\Big)-\frac{m^2\lambda_{12}}{4}\bigg\}.\notag
\end{align}
Due to the $q_1\leftrightarrow -q_2$ symmetry of the integrand, the $t$-
and $u$-channel terms give the same contribution. 

We stress that in our dispersive framework the analytic structure of the
HLbL tensor has to be analyzed for the full four-point function, i.e.\ with
$k^2=0$ but otherwise general $k$. In this setting $s$ and $q_3^2$ are
independent variables and the pion-pole contribution to the HLbL
tensor is unambiguously given by~\eqref{lbl_pi0pole}, which leads to the result
in~\eqref{amu_pole}. 
Within our formalism, the pion pole defined in this manner~\eqref{amu_pole} is
therefore unique. 

In~\cite{MV} it was pointed out that the pion-pole contribution as defined
in~\eqref{lbl_pi0pole} goes faster to zero for large $q^2$ than what is
required by perturbative QCD, thereby becoming sub-dominant in that regime.
As a cure to this problem, it was proposed in~\cite{MV} to replace the
singly-virtual form factor by a constant, arguing that in this way one
obtains an expression which correctly interpolates between high and low
$q^2$. As stated in that paper, the transition
$\F_{\pi^0\gamma^*\gamma^*}(q^2,0) \to {\it const}$ for large $q^2$ is
generated by the exchange of heavier pseudoscalar resonances, which we are
explicitly neglecting here. We are well aware that restoring the correct
high-energy behavior has a non-negligible impact in the numerical estimate
of the HLbL contribution to $a_\mu$. Therefore, enforcing additional
short-distance constraints onto our representation is indeed a direction
for future improvements of the formalism~\cite{CHPS_prep}.
This statement pertains not only to the pion pole, but to a dispersive
approach in general: with a limited number of intermediate states explicitly 
taken into account, the representation will be adequate at low and intermediate
energies, while the correct high-energy behavior has to be enforced
in a second step.
 
We also observe that the leading and subleading logarithmic contributions
to $a_\mu$ in a chiral and $1/N_C$ expansion discussed in~\cite{RamseyMusolf:2002cy} are automatically reproduced in this approach:
the leading one by construction, and the subleading one provided the
measured decays $\eta \to \mu^+ \mu^-$ and $\pi^0 \to e^+ e^-$ are used to
constrain the off-shell dependence in
$\F_{\pi^0\gamma^*\gamma^*}\big(q_1^2,q_2^2\big)$.

We conclude with a brief comment about an alternative, model-dependent
implementation of the pole amplitude in which the pion transition form
factor is generalized to arbitrary pion virtualities~\cite{JN}, e.g.\
$\F_{\pi^0\gamma^*\gamma^*}(s,s,0)$ for the $s$-channel pole (the first
argument referring to the pion virtuality). When expanded around the pole
mass, any offshell-dependence from the form factors yields a polynomial
contribution, as long as it does not entail more complicated analytic
structure (and thus intermediate states beyond our framework). In the
dispersive picture a polynomial arises if the dispersive integrals do not
converge sufficiently fast and have to be subtracted. Off-shell terms
calculated in a given model can be absorbed into this polynomial.
Frequently, the coefficients of the subtraction polynomial are free
parameters in a dispersive approach, but for HLbL scattering constraints by
gauge invariance
completely fix these parameters, as will be shown in Sects.~\ref{sec:disp_rel}
and~\ref{sec:non_diagonal}.

\subsection{Conventions for $\boldsymbol{\gamma^*\gamma^*\to\pi\pi}$}
\label{sec:ggpipi}

In this section we collect notation and conventions for
$\gamma^*\gamma^*\to\pi\pi$. We take pion Compton scattering  
\beq
\gamma^*(q_1,\lambda_1,\mu)\pi^a(p_1)\to\gamma^*(q_2,\lambda_2,\nu)\pi^b(p_2),
\eeq
with momenta, helicities, and Lorentz indices as indicated as well as
isospin indices $a$, $b$ as the $s$-channel process and define Mandelstam
variables according to\footnote{To keep the notation as simple as possible
  we use the same symbols as before, the understanding being that in this
  section $s$, $t$, etc.\ refer to $\gamma^*\pi\to\gamma^*\pi$ kinematics,
  but in the rest of the paper to HLbL scattering.} 
\beq
s=(q_1+p_1)^2,\qquad t=(q_1-q_2)^2,\qquad u=(q_1-p_2)^2.
\eeq
The scattering angle in the crossed channel
\beq
\label{ggpipi}
\gamma^*(q_1,\lambda_1,\mu)\gamma^*(-q_2,\lambda_2,\nu)\to\pi^a(-p_1)\pi^b(p_2)
\eeq
is given by
\beq
z_t=\frac{s-u}{4p_tq_t},
\eeq
with momenta
\beq
p_t=\sqrt{\frac{t}{4}-\mpi^2}\equiv\frac{\sqrt{t}}{2}\sigma_t,\qquad q_t=\frac{\sqrt{\lambda_{12}^t}}{2\sqrt{t}}, \qquad
\lambda_{12}^t=\lambda\big(t,q_1^2,q_2^2\big).
\eeq

The objects of interest for HLbL scattering are the helicity amplitudes corresponding to~\eqref{ggpipi}
\begin{align}
{}_\text{out}\langle \pi(-p_1)\pi(p_2)|\gamma^*(q_1,\lambda_1)\gamma^*(-q_2,\lambda_2)\rangle_\text{in}
&=ie^2(2\pi)^4\delta^4(q_1+p_1-q_2-p_2)H_{\lambda_1\lambda_2}e^{i(\lambda_1-\lambda_2)\varphi}\notag\\
H_{\lambda_1\lambda_2}&=\eps_\mu(q_1,\lambda_1)\eps_\nu(-q_2,\lambda_2)W^{\mu\nu},
\end{align}
parameterized in terms of the tensor $W^{\mu\nu}$ ($\varphi$ denotes the azimuthal angle). Its Lorentz decomposition into gauge-invariant structures that separately fulfill the Ward identities
\beq
q_1^\mu W_{\mu\nu}= q_2^\nu W_{\mu\nu}=0
\eeq
reads
\beq
 W_{\mu\nu}=\sum_{i=1}^5T_{\mu\nu}^iA_i,
 \eeq
 where
 \begin{align}
 \label{basis}
 T^1_{\mu\nu}&=-q_1\cdot q_2 g_{\mu\nu}+ q_{2\mu} q_{1\nu},\notag\\
 T^2_{\mu\nu}&=-4q_1\cdot q_2\Delta_\mu\Delta_\nu-(s-u)^2g_{\mu\nu}+2(s-u)\big(\Delta_\mu q_{1\nu}+q_{2\mu}\Delta_\nu \big),\notag\\
 T^3_{\mu\nu}&=q_1\cdot q_2q_{1\mu}q_{2\nu}+q_1^2q_2^2g_{\mu\nu}-q_2^2q_{1\mu}q_{1\nu}-q_1^2q_{2\mu}q_{2\nu},\notag\\
 T^4_{\mu\nu}&=-2q_1\cdot q_2q_{1\mu}\Delta_\nu+2q_1^2q_{2\mu}\Delta_\nu-(s-u)q_1^2g_{\mu\nu}+(s-u)q_{1\mu}q_{1\nu},\notag\\
 T^5_{\mu\nu}&=-2q_1\cdot q_2\Delta_\mu q_{2\nu}+2q_2^2\Delta_\mu q_{1\nu}-(s-u)q_2^2g_{\mu\nu}+(s-u)q_{2\mu}q_{2\nu},
\end{align}
and
\beq
\Delta_\mu=p_{1\mu}+p_{2\mu}.
\eeq

To work out the explicit form of the helicity amplitudes, we choose momenta and polarization vectors as follows\footnote{We choose the signs of the transversal helicities in accordance with the conventions in~\cite{Edmonds}.}
\begin{align}
 q_1^\mu&=(E_{q_1},0,0,q_t),\qquad  q_2^\mu=(-E_{q_2},0,0,q_t),\notag\\
 p_1^\mu&=(-E_{p_1},-p_t\sin\theta_t,0,-p_t\cos\theta_t),\qquad 
 p_2^\mu=(E_{p_2},-p_t\sin\theta_t,0,-p_t\cos\theta_t),\notag\\
 \eps^\mu(q_1,\pm)&=\mp\frac{1}{\sqrt{2}}(0,1,\pm i,0),\qquad \eps^\mu(-q_2,\pm)=\mp\frac{1}{\sqrt{2}}(0,1,\mp i,0),\notag\\
 \eps^\mu(q_1,0)&=\frac{1}{\xi_1}(q_t,0,0,E_{q_1}),\qquad
 \eps^\mu(-q_2,0)=\frac{1}{\xi_2}(-q_t,0,0,E_{q_2}),\notag\\
 E_{q_1}&=\frac{t+q_1^2-q_2^2}{2\sqrt{t}},\qquad 
 E_{q_2}=\frac{t-q_1^2+q_2^2}{2\sqrt{t}},\qquad
 E_{p_1}=E_{p_2}=\frac{\sqrt{t}}{2},
\end{align}
leaving the normalization $\xi_i$ of the longitudinal polarization vectors
general. In these conventions, we find the following expressions for the
helicity amplitudes\footnote{Bose symmetry in the pion system requires that
  the full amplitude remain invariant under $p_1 \leftrightarrow
  -p_2$. Since $T^{4}_{\mu\nu}$ and $T^{5}_{\mu\nu}$ are odd under this
  transformation, the corresponding scalar functions need to be odd as
  well, so that one factor in $s-u$ can be separated without introducing
  kinematic singularities. Accordingly, $A_4$ and $A_5$ are eliminated, in
  practice, in favor of $\tilde A_4=A_4/(s-u)$ and $\tilde A_5=A_5/(s-u)$.} 
\begin{align}
\label{helicity_amplitudes}
 H_{++}=H_{--}&=-\frac{1}{2}\big(t-q_1^2-q_2^2\big)A_1+\big(t-4\mpi^2\big)\Big\{\big(t-q_1^2-q_2^2\big)\big(1-z_t^2\big)+4q_t^2z_t^2\Big\}A_2
 -q_1^2q_2^2A_3\notag\\&+4p_tq_tz_t\Big(q_1^2A_4+q_2^2A_5\Big),\notag\\
 H_{+-}=H_{-+}&=-\big(t-4\mpi^2\big)\big(t-q_1^2-q_2^2\big)\big(1-z_t^2\big)A_2,\notag\\
 H_{+0}=-H_{-0}&=\frac{q_2^2}{\xi_2}\bigg[-\big(t-4\mpi^2\big)\sqrt{\frac{2}{t}}\big(t+q_1^2-q_2^2\big)z_t\sqrt{1-z_t^2}A_2
 +2p_tq_t\sqrt{2t}\sqrt{1-z_t^2}A_5\bigg],\notag\\
 H_{0+}=-H_{0-}&=\frac{q_1^2}{\xi_1}\bigg[-\big(t-4\mpi^2\big)\sqrt{\frac{2}{t}}\big(t-q_1^2+q_2^2\big)z_t\sqrt{1-z_t^2}A_2
 +2p_tq_t\sqrt{2t}\sqrt{1-z_t^2}A_4\bigg],\notag\\
 H_{00}&=-\frac{q_1^2q_2^2}{\xi_1\xi_2}\Big\{A_1-4\big(t-4\mpi^2\big)z_t^2 A_2+\frac{1}{2}\big(t-q_1^2-q_2^2\big)A_3
 +4p_tq_tz_t\big(A_4+A_5\big)\Big\}.
\end{align}
We define the partial waves as
\begin{align}
\label{def_partial_waves}
 H_{++}(s,t)&=\sum\limits_{J=0}^\infty(2J+1)P_J(z_t)h_{J,++}(t),\notag\\
 H_{+-}(s,t)&=\sum\limits_{J=2}^\infty(2J+1)d^J_{20}(z_t)h_{J,+-}(t),\notag\\
 H_{+0}(s,t)&=\sum\limits_{J=2}^\infty(2J+1)d_{10}^J(z_t)h_{J,+0}(t),\notag\\
 H_{0+}(s,t)&=\sum\limits_{J=2}^\infty(2J+1)d_{10}^J(z_t)h_{J,0+}(t),\notag\\
 H_{00}(s,t)&=\sum\limits_{J=0}^\infty(2J+1)P_J(z_t)h_{J,00}(t),
\end{align}
where $P_J(z)$ denotes Legendre polynomials and $d^J_{mm'}(z)$ Wigner $d$-functions, e.g.\
\beq
d^2_{20}(z)=\frac{\sqrt{6}}{4}\big(1-z^2\big),\qquad
d^2_{10}(z)=-\sqrt{\frac{3}{2}}z\sqrt{1-z^2}. 
\eeq
Due to Bose symmetry and invariance of strong interactions under charge
conjugation, only even partial waves are allowed to contribute. 
$S$-waves only occur for the $++$ and $00$ projection, while all other
helicity projections start at $D$-wave level. Explicit expressions for
the partial-wave projections of the Born terms, $N_{J,\lambda_1\lambda_2}$,
are given in App.~\ref{app:FsQED}.

The helicity partial waves for $\gamma^*\gamma^*\to\pi\pi$ as defined
in~\eqref{def_partial_waves} constitute the key input for $\pi\pi$
intermediate states in the HLbL contribution to $a_\mu$. They fulfill
important constraints known as soft-photon zeros~\cite{Low,Moussallam13},
which will prove crucial for the construction of dispersion relations in
Sect.~\ref{sec:disp_rel}. The soft-photon theorem states that 
in the limit $q_1^2\to 0$ the Born-term-subtracted helicity amplitudes vanish if $t=q_2^2$ and vice versa.
Its origin can be inferred from the
decomposition~\eqref{helicity_amplitudes}.  For later use it
will be convenient to express these soft-photon zeros more explicitly on
the level of partial waves, in the framework of partial-wave dispersion
relations (see~\cite{GMM,HPS} for the on-shell case $\gamma\gamma\to\pi\pi$
and~\cite{Moussallam13} for a generalization to the singly-virtual process
$\gamma^*\gamma\to\pi\pi$). In particular, one can derive a system of
so-called Roy--Steiner equations  
\beq
\label{RSsystem}
h_{J,i}(t)=\frac{1}{\pi}\sum\limits_{J'\,\text{even}}\sum\limits_{j=1}^5\int\limits_{4\mpi^2}^\infty\diff t' K_{JJ'}^{ij}(t,t')\Im h_{J',j}(t')+\cdots,\qquad i,j\in\big\{++,+-,+0,0+,00\},
\eeq
where the ellipsis refers to integrals involving partial waves for $\gamma^*\pi\to\gamma^*\pi$~\cite{HPS,CHPS_prep}. The precise form how the soft-photon zeros manifest themselves may be read off from the diagonal kernel functions
\begin{align}
\label{kernels_diagonal}
 K_{00}^{11}(t,t')&=K_{00}^{55}(t,t')=\frac{1}{t'-t}-\frac{t'-q_1^2-q_2^2}{\lambda_{12}^{t'}},\notag\\
 K_{22}^{11}(t,t')&=K_{22}^{55}(t,t')=\frac{p_t^2q_t^2}{p_t'^2q_t'^2}\bigg(\frac{1}{t'-t}-\frac{t'-q_1^2-q_2^2}{\lambda_{12}^{t'}}\bigg),\notag\\
 K_{22}^{22}(t,t')&=\frac{\big(t-q_1^2-q_2^2\big) p_t^2}{(t'-t) \big(t'-q_1^2-q_2^2\big) p_t'^2},\notag\\
 K_{22}^{33}(t,t')&=K_{22}^{44}(t,t')=\sqrt{\frac{t}{t'}}\frac{p_t^2 q_t^2}{(t'-t) p_t'^2 q_t'^2}.
\end{align}
The form of these kernel functions will be instrumental for the construction of dispersion relations for HLbL scattering in Sect.~\ref{sec:disp_rel}.
Apart from the diagonal kernel functions~\eqref{kernels_diagonal} there are also non-diagonal ones, e.g.\ for the $S$-waves
\beq
\label{kernels_non_diagonal}
K_{00}^{15}(t,t')=\frac{2\xi_1\xi_2}{\lambda_{12}^{t'}},\qquad
K_{00}^{51}(t,t')=\frac{2q_1^2q_2^2}{\xi_1\xi_2\lambda_{12}^{t'}}.
\eeq
The necessity of these additional kernel functions using the example of the $1$-loop amplitudes in Chiral Perturbation Theory (ChPT) is demonstrated in App.~\ref{app:ggpipi_ChPT}.

\section{$\boldsymbol{\pi\pi}$ intermediate states}
\label{sec:master}

In this section we discuss a dispersive treatment of the $\bar
\Pi_{\mu\nu\lambda\sigma}$ part of the HLbL tensor defined in
\eqref{eq:Pibreakdown}. 
Modifications due to the subtraction of the FsQED loop, as
well as the symmetry factor for $\pi^0\pi^0$ intermediate states, will be
discussed in more detail in Sect.~\ref{subsec:master}.

The outline of the derivation is as follows: in Sect.~\ref{sec:unitarity_decomp} we first analyze the unitarity relation for $\pi\pi$ intermediate states. This leads to a set of Lorentz structures diagonal in the space of helicity amplitudes that serves as a basis for the HLbL tensor. Dispersion relations are then written down for the scalar coefficients of these Lorentz structures. The form of the diagonal kernel functions of these dispersion relations can be immediately read off from $\gamma^*\gamma^*\to\pi\pi$, as detailed in Sect.~\ref{sec:disp_rel}.
Apart from the diagonal kernels, there are in general non-diagonal contributions, in analogy to~\eqref{kernels_non_diagonal}. In Sect.~\ref{sec:non_diagonal} we discuss the origin of these terms and derive their explicit form for $S$-waves, before presenting our main result
in Sect.~\ref{subsec:master}.

\subsection{Unitarity and decomposition of the HLbL amplitude}
\label{sec:unitarity_decomp}

\begin{figure}
\centering
\includegraphics[width=0.45\linewidth]{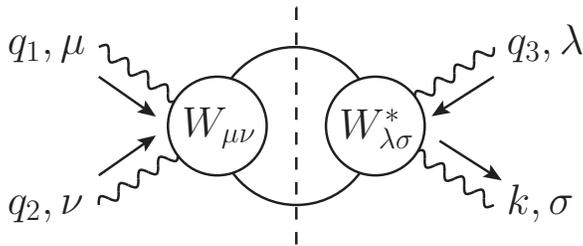}
\caption{Unitarity relation for $\pi\pi$ intermediate states in HLbL
  scattering. The blobs refer to the full $\gamma^*\gamma^*\to\pi\pi$ amplitude, otherwise conventions
   as in Fig.~\ref{fig:sQED}.}
\label{fig:pipi_unitarity}
\end{figure}

We start off from the unitarity relation for $\pi\pi$ intermediate states, as illustrated in Fig.~\ref{fig:pipi_unitarity}.
The $s$-channel discontinuity of $\Pi^{\mu\nu\lambda\sigma}$ due to $\pi\pi$ intermediate states follows from
\begin{align}
\label{pipiunitarity}
\Imspipi \Pi_{\mu\nu\lambda\sigma}&=\frac{1}{2}\int \frac{\diff^3k_1}{(2\pi)^32E_1}\int \frac{\diff^3k_2}{(2\pi)^32E_2}
W_{\mu\nu}\big(\gamma^*(q_1,\mu)\gamma^*(q_2,\nu)\to\pi(k_1)\pi(k_2)\big)\notag\\
&\times W^*_{\lambda\sigma}\big(\gamma^*(q_3,\lambda)\gamma(-k,\sigma)\to\pi(-k_1)\pi(-k_2)\big)
(2\pi)^4\delta^4\big(k_1+k_2-q_1-q_2\big).
\end{align}
The phase-space integral in~\eqref{pipiunitarity} may be rewritten in terms of loop integrals
\begin{align}
\Imspipi \Pi^{\mu\nu\lambda\sigma}&=\Ims\frac{1}{i}\int \frac{\diff^4l}{(2\pi)^4}
\frac{
1}{\big(l^2-\mpi^2\big)\big((l-\Sigma)^2-\mpi^2\big)}\\
&\times 
W_{\mu\nu}\big(\gamma^*(q_1,\mu)\gamma^*(q_2,\nu)\to\pi(l)\pi(\Sigma-l)\big) W^*_{\lambda\sigma}\big(\gamma^*(q_3,\lambda)\gamma(-k,\sigma)\to\pi(-l)\pi(l-\Sigma)\big),\notag
\end{align}
where $\Sigma=q_1+q_2$. This relation can be analyzed at a given order in a partial-wave expansion for $\gamma^*\gamma^*\to\pi\pi$ in terms of tensor integrals. Including partial waves up to $D$-waves, we find
\begin{align}
\label{Imsresult}
 \Imspipi \Pi^{\mu\nu\lambda\sigma}&=\frac{\sigma_s}{16\pi}\theta\big(s-4\mpi^2\big)\bigg\{
 \Big(h_1\big(s;q_1^2,q_2^2\big)h_1^*\big(s;q_3^2,0\big)+\frac{1}{5}P_2(z_s)h_3\big(s;q_1^2,q_2^2\big)h_3^*\big(s;q_3^2,0\big)\Big)A_{1,s}^{\mu\nu\lambda\sigma}\notag\\
 &+\Big(h_2\big(s;q_1^2,q_2^2\big)h_1^*\big(s;q_3^2,0\big)+\frac{1}{5}P_2(z_s)h_7\big(s;q_1^2,q_2^2\big)h_3^*\big(s;q_3^2,0\big)\Big)A_{2,s}^{\mu\nu\lambda\sigma}\notag\\
 &+h_3\big(s;q_1^2,q_2^2\big)\Big(h_4^*\big(s;q_3^2,0\big)A_{3,s}^{\mu\nu\lambda\sigma}+h_6^*\big(s;q_3^2,0\big)A_{4,s}^{\mu\nu\lambda\sigma}\Big)\notag\\
 &+h_4\big(s;q_1^2,q_2^2\big)\Big(h_3^*\big(s;q_3^2,0\big)A_{5,s}^{\mu\nu\lambda\sigma}+h_4^*\big(s;q_3^2,0\big)A_{6,s}^{\mu\nu\lambda\sigma}+h_6^*\big(s;q_3^2,0\big)A_{7,s}^{\mu\nu\lambda\sigma}\Big)\notag\\
 &+h_5\big(s;q_1^2,q_2^2\big)\Big(h_3^*\big(s;q_3^2,0\big)A_{8,s}^{\mu\nu\lambda\sigma}+h_4^*\big(s;q_3^2,0\big)A_{9,s}^{\mu\nu\lambda\sigma}+h_6^*\big(s;q_3^2,0\big)A_{10,s}^{\mu\nu\lambda\sigma}\Big)\notag\\
 &+h_6\big(s;q_1^2,q_2^2\big)\Big(h_3^*\big(s;q_3^2,0\big)A_{11,s}^{\mu\nu\lambda\sigma}+h_4^*\big(s;q_3^2,0\big)A_{12,s}^{\mu\nu\lambda\sigma}+h_6^*\big(s;q_3^2,0\big)A_{13,s}^{\mu\nu\lambda\sigma}\Big)\notag\\
 &+h_7\big(s;q_1^2,q_2^2\big)\Big(h_4^*\big(s;q_3^2,0\big)A_{14,s}^{\mu\nu\lambda\sigma}+h_6^*\big(s;q_3^2,0\big)A_{15,s}^{\mu\nu\lambda\sigma}\Big)
 \bigg\},
 \end{align}
 with
 \begin{align}
 h_1\big(s;q_1^2,q_2^2\big)&=h_{0,++}\big(s;q_1^2,q_2^2\big), 
& h_5\big(s;q_1^2,q_2^2\big)&=5\sqrt{\frac{3}{2}}\frac{\xi_2}{q_2^2}\sqrt{\frac{s}{2}}h_{2,+0}\big(s;q_1^2,q_2^2\big),\notag\\
 h_2\big(s;q_1^2,q_2^2\big)&=-\frac{\xi_1\xi_2}{q_1^2q_2^2}h_{0,00}\big(s;q_1^2,q_2^2\big), 
& h_6\big(s;q_1^2,q_2^2\big)&=5\sqrt{\frac{3}{2}}\frac{\xi_1}{q_1^2}\sqrt{\frac{s}{2}}h_{2,0+}\big(s;q_1^2,q_2^2\big),\notag\\
 h_3\big(s;q_1^2,q_2^2\big)&=5h_{2,++}\big(s;q_1^2,q_2^2\big),
&  h_7\big(s;q_1^2,q_2^2\big)&=-5\frac{\xi_1\xi_2}{q_1^2q_2^2}h_{2,00}\big(s;q_1^2,q_2^2\big),\notag\\
 h_4\big(s;q_1^2,q_2^2\big)&=-\frac{5\sqrt{6}}{4}h_{2,+-}\big(s;q_1^2,q_2^2\big),
\end{align}
indicating the photon virtualities in the argument of the partial waves, and Lorentz structures $A_{i,s}^{\mu\nu\lambda\sigma}$ as summarized in App.~\ref{app:Lorentz}. The discontinuities for $t$- and $u$-channel involve the tensors $A_{i,t}^{\mu\nu\lambda\sigma}$
and $A_{i,u}^{\mu\nu\lambda\sigma}$, which
follow from the $s$-channel analysis by means of crossing symmetry
$(q_2,\nu) \leftrightarrow (q_3,\lambda)$ and $(q_1,\mu) \leftrightarrow
(q_3,\lambda)$, respectively. The fifteen Lorentz structures which have
emerged from the unitarity relation represent a key element for our
derivation of the dispersion relation for the HLbL tensor, as we are now
going to explain.

In general, the HLbL tensor with one of the four photons on-shell 
contains $29$ independent scalar amplitudes. We have explicitly constructed
$29$ independent gauge-invariant Lorentz tensors, but doing so in a way
that makes crossing symmetry manifest, or even easy to express, is more
difficult. For our purposes we find it more convenient to use a redundant
basis, in which however crossing symmetry is evident. Therefore, we exploit
the crucial property of the $A_{i,s}^{\mu\nu\lambda\sigma}$ that if we
project the $s$-channel HLbL tensor on helicity amplitudes, only a single
function $\Pi_i^s\equiv\Pi_i(s,t,u)$ contributes for each helicity
amplitude, and write 
\beq
\label{lbl_tensor}
\bar \Pi^{\mu\nu\lambda\sigma}(s,t,u)=\sum_{i=1}^{15}\Big(A_{i,s}^{\mu\nu\lambda\sigma}\Pi_i(s,t,u)
+A_{i,t}^{\mu\nu\lambda\sigma}\Pi_i(t,s,u)+A_{i,u}^{\mu\nu\lambda\sigma}\Pi_i(u,t,s)\Big).
\eeq
We have checked that the $45$ tensors
in~\eqref{lbl_tensor} form a complete, though redundant, basis. In fact,
already the $30$ tensors $A_{i,s}^{\mu\nu\lambda\sigma}$ and 
$A_{i,t}^{\mu\nu\lambda\sigma}$ alone saturate the number of linearly
independent structures, with just one redundant tensor.
If we project the whole tensor onto $s$-channel helicity amplitudes, besides the diagonal contribution from $\Pi_i^s$, we
will get contributions from all $\Pi_i^t$ and $\Pi_i^u$ to each helicity
amplitude. Explicitly, the first few $s$-channel helicity amplitudes
 as defined in~\eqref{eq:H1234}
read\footnote{The bar over the helicity amplitudes $\bar \Pi_{\lambda_1\lambda_2,\lambda_3\lambda_4}$ indicates that these are the projections of the tensor $\bar \Pi^{\mu\nu\lambda\sigma}$.}
\begin{align}
\label{eq:HelPi}
\bar H_{++,++}(s,t,u) &= \Pi_1(s,t,u)+\hat{H}_{++,++}(s,t,u),\notag \\
\bar H_{00,++}(s,t,u) &=-\frac{q_1^2q_2^2}{\xi_1 \xi _2} \Pi_2(s,t,u)+\hat{H}_{00,++}(s,t,u),\notag\\
\bar H_{++,+-}(s,t,u) &=-\frac{4}{5 \sqrt{6}}d^2_{02}(z_s) \Pi_3(s,t,u)+\hat{H}_{++,+-}(s,t,u),
\end{align}
and similarly for the remaining ones. The hat amplitudes defined by 
\beq
\hat{H}_{\lambda_1 \lambda_2,\lambda_3 \lambda_4}(s,t,u) = \sum_{i=1}^{15}
\left( f_{\lambda_1 \lambda_2,\lambda_3 \lambda_4}^i
  \Pi_i(t,s,u)+\tilde{f}_{\lambda_1 \lambda_2,\lambda_3 \lambda_4}^i
  \Pi_i(u,t,s) \right)
\eeq
with coefficients $f^i$ and $\tilde{f}^i$ obtained from the contraction of
the polarization vectors with $A_{i,t}^{\mu\nu\lambda\sigma}$ and
$A_{i,u}^{\mu\nu\lambda\sigma}$, respectively, are responsible for the
left-hand cut in the corresponding helicity amplitude. The right-hand cut
for $s \geq 4 \mpi^2$ on the other hand is 
solely incorporated in the $\Pi_i^s$ amplitude. 
In this sense, the decomposition~\eqref{lbl_tensor} leads to diagonal
unitarity relations for the helicity amplitudes, see App.~\ref{app:unitarity_hel}.  

The final step in the dispersive calculation of $\bar \Pi^{\mu\nu\lambda\sigma}$
concerns the construction of dispersion relations for the coefficient
functions $\Pi_i$. In the next section, we determine the diagonal kernel functions by comparison with
$\gamma^*\gamma^*\to\pi\pi$, while the issue of non-diagonal kernels will be discussed in Sect.~\ref{sec:non_diagonal}.

\subsection{Dispersion relations: diagonal kernel functions}
\label{sec:disp_rel}

The construction of dispersion relations for the $\Pi_i$ functions
becomes greatly simplified if we consider that here we are only
interested in the HLbL contribution to $a_\mu$. This involves the
derivative of the HLbL tensor with respect to $k$ evaluated at $k=0$. We
therefore construct dispersion relations only for this very special limit
and omit from the start any contribution to the HLbL tensor of
$\Order(k^2)$. Since the $A_{i,s}^{\mu\nu\lambda\sigma}$ scale as  
$\Order(k^0)$, any contribution of $\Order(k^2)$ in $\Pi_i$ can be 
dropped right away. In particular, this concerns most of the $D$-wave
contributions due to the angular-momentum factor\footnote{We discuss the
  case of $D$-waves in the $s$-channel here, but the same argument applies
  to $D$-waves in all other channels.} 
\beq
\label{threshold_Dwave}
q_{34}^2=\frac{\big(s-q_3^2\big)^2}{4s}=\Order\big(k^2\big),
\eeq
whereas for $S$-waves the expected overall $\Order(k)$ scaling is restored
by the soft-photon zero, which requires a factor 
\beq
s-q_3^2=-2k\cdot q_3+\Order\big(k^2\big).
\eeq
Not all $D$-wave contributions to $a_\mu$ vanish, however. As it follows
from the decomposition of the $\gamma^*\gamma^*\to\pi\pi$ helicity
amplitudes in~\eqref{helicity_amplitudes}, or, alternatively, the kernel
function $K_{22}^{22}(t,t')$ in~\eqref{kernels_diagonal},\footnote{In the
  present   context, one should take $q_1^2\to q_3^2$, $q_2^2\to0$, and
  $t\to s$ in these equations.} the threshold behavior for the $+-$
system differs from~\eqref{threshold_Dwave}, in the sense that only a
single factor in $s-q_3^2$ appears. Such $D$-waves do contribute to $a_\mu$
since after taking the derivative with respect to $k$ and the $k\to
0$ limit they do not vanish. Ultimately, this special threshold behavior
is a consequence of gauge invariance,
which dictates the general decomposition in~\eqref{helicity_amplitudes}.  

Considering all terms in~\eqref{lbl_tensor} we find that the only diagonal kernels
that contribute in the end are those with
$i\in\{1,2,3,6,9,12,14\}$. Moreover, for $i\in\{1,2\}$ only $S$-waves are
relevant, with the $D$-waves exhibiting the angular-momentum factor as
in~\eqref{threshold_Dwave}, while the other non-vanishing terms, $i\in\{3,6,9,12,14\}$, start at
$D$-wave level in the first place. Since even there the dependence on the
scattering angle is completely hidden in $A_{i,s}^{\mu\nu\lambda\sigma}$
(or its crossed versions), this implies that all $\Pi_i$ effectively become
single-variable functions, which, by construction, only have a right-hand
cut. Unitarity fixes the discontinuity of these functions, see
App.~\ref{app:unitarity_hel}, on the right-hand cut and therefore the whole
function, up to a polynomial contribution. As we will show, however, this
polynomial is completely fixed by soft-photon constraints.  We stress that
the separation of right- and left-hand cut---the property of the $\Pi_i$
amplitudes to only have a right-hand cut---is possible only in the absence
of double-spectral regions, which holds true for $\pi\pi$ intermediate
states after separating the FsQED pion loop as discussed in
Sect.~\ref{sec:disprel}. This reduction to single-variable dispersion
relations can also be understood in the framework of $\pi\pi$ Roy
equations, where it occurs if one neglects the contribution to the
discontinuities from non-leading partial waves, as explained in
App.~\ref{app:roy}. The original idea, which goes under the name of
reconstruction theorem, has been first formulated in~\cite{Stern:1993rg}.

Based on the discussion above, we construct a dispersive representation of the $\Pi_i^s$
amplitudes which has the following properties
\begin{enumerate}
\item
For each $\Pi_i^s$ we only take into account the discontinuity due to the
lowest partial wave.
\item
We fix the discontinuity to what unitarity prescribes.
\item
The $\Pi_i^s$ amplitudes have the required soft-photon zeros.
\item
The exact form of the soft-photon zeros follows from $\gamma^*\gamma^*\to \pi\pi$ by means of factorization. 
\item
The number of subtractions is chosen according to what the implementation of the
soft-photon zeros naturally generates (which is sufficient to ensure convergence).
\end{enumerate}
Soft-photon zeros are also properties of the $\gamma^* \gamma^* \to \pi
\pi $ sub-amplitudes, where they manifest themselves as a modification of
the Cauchy kernel. The form of the kernel functions of the
dispersion relations for the $\Pi_i^s$ can be read off from the
$\gamma^*\gamma^*\to\pi\pi$ kernel functions in~\eqref{kernels_diagonal} in
the following way. After dropping the pion angular-momentum factors, these
kernel functions lead to modifications of the Cauchy kernel due to
soft-photon zeros and photon angular-momentum factors 
\beq
K_{12}(s,s')=\frac{f_{12}(s,s')}{s'-s},\qquad K_{34}(s,s')=\frac{f_{34}(s,s')}{s'-s},
\eeq
for the initial- and final-state photon pair, respectively. The
corresponding kernel in HLbL scattering, which has exactly the same
soft-photon behavior in each sub-amplitude, can be easily obtained by
factorization
\beq
K_{12,34}(s,s')=\frac{f_{12}(s,s')f_{34}(s,s')}{s'-s}.
\eeq
These arguments uniquely lead to the following dispersive integrals for
the $\Pi_i^s$ amplitudes\footnote{This representation neglects non-diagonal terms, which will be discussed in Sect.~\ref{sec:non_diagonal}.}
\begin{align}
\label{disp_rel_diag}
 \Pi_1^s&=\bar h^0_{++,++}(s)=\frac{s-q_3^2}{\pi}\int\limits_{4\mpi^2}^\infty\frac{\diff s'}{s'-q_3^2}\bigg(\frac{1}{s'-s}
 -\frac{s'-q_1^2-q_2^2}{\lambda_{12}'}\bigg)\Im \bar h_{++,++}^0(s'),\\
  -\frac{q_1^2q_2^2}{\xi_1\xi_2}\Pi_2^s&=\bar h^0_{00,++}(s)=\frac{s-q_3^2}{\pi}\int\limits_{4\mpi^2}^\infty\frac{\diff s'}{s'-q_3^2}
  \bigg(\frac{1}{s'-s}-\frac{s'-q_1^2-q_2^2}{\lambda_{12}'}\bigg)\Im \bar h_{00,++}^0(s'),\notag\\
-\frac{2\sqrt{6}}{75}\Pi_3^s&=\bar h^2_{++,+-}(s)=\frac{\big(s-q_3^2\big)\lambda_{12}}{s\,\pi}\int\limits_{4\mpi^2}^\infty\frac{\diff s'\, s'}{\big(s'-q_3^2\big)\lambda_{12}'}
  \bigg(\frac{1}{s'-s}-\frac{s'-q_1^2-q_2^2}{\lambda_{12}'}\bigg)\Im \bar h_{++,+-}^2(s'),\notag\\
  \frac{8}{75}\Pi_6^s&=\bar h^2_{+-,+-}(s)=\frac{\big(s-q_3^2\big)\big(s-q_1^2-q_2^2\big)}{\pi}\int\limits_{4\mpi^2}^\infty\diff s'
  \frac{\Im \bar h_{+-,+-}^2(s')}{\big(s'-q_3^2\big)\big(s'-q_1^2-q_2^2\big)\big(s'-s\big)},\notag\\
-\frac{4}{75}\sqrt{\frac{2}{s}}\frac{q_2^2}{\xi_2}\Pi_9^s&=\bar h^2_{+0,+-}(s)=\frac{\big(s-q_3^2\big)\lambda_{12}}{\sqrt{s}\,\pi}\int\limits_{4\mpi^2}^\infty\diff s'
  \frac{\sqrt{s'}\,\Im \bar h_{+0,+-}^2(s')}{\big(s'-q_3^2\big)\lambda_{12}'\big(s'-s\big)},\notag\\
-\frac{4}{75}\sqrt{\frac{2}{s}}\frac{q_1^2}{\xi_1}\Pi_{12}^s&=\bar h^2_{0+,+-}(s)=\frac{\big(s-q_3^2\big)\lambda_{12}}{\sqrt{s}\,\pi}\int\limits_{4\mpi^2}^\infty\diff s'
  \frac{\sqrt{s'}\,\Im \bar h_{0+,+-}^2(s')}{\big(s'-q_3^2\big)\lambda_{12}'\big(s'-s\big)},\notag\\
\frac{2\sqrt{6}}{75}\frac{q_1^2q_2^2}{\xi_1\xi_2}\Pi_{14}^s&=\bar h^2_{00,+-}(s)=\frac{\big(s-q_3^2\big)\lambda_{12}}{s\,\pi}\int\limits_{4\mpi^2}^\infty\frac{\diff s'\, s'}{\big(s'-q_3^2\big)\lambda_{12}'}
  \bigg(\frac{1}{s'-s}-\frac{s'-q_1^2-q_2^2}{\lambda_{12}'}\bigg)\Im \bar h_{00,+-}^2(s'),\notag
\end{align}
and accordingly for the crossed channels, with imaginary parts
\begin{align}
\label{imaginary_parts}
 \Im \bar h_{++,++}^0(s)&=\frac{\sigma_s}{16\pi}\theta\big(s-4\mpi^2\big)\,\Sym\Big[h_{0,++}\big(s;q_1^2,q_2^2\big)h_{0,++}^*\big(s;q_3^2,0\big)\Big],\notag\\
 \Im \bar h_{00,++}^0(s)&=\frac{\sigma_s}{16\pi}\theta\big(s-4\mpi^2\big)\,\Sym\Big[h_{0,00}\big(s;q_1^2,q_2^2\big)h_{0,++}^*\big(s;q_3^2,0\big)\Big],\notag\\
 \Im \bar h_{++,+-}^2(s)&=\frac{\sigma_s}{16\pi}\theta\big(s-4\mpi^2\big)\,\Sym\Big[h_{2,++}\big(s;q_1^2,q_2^2\big)h_{2,+-}^*\big(s;q_3^2,0\big)\Big],\notag\\
 \Im \bar h_{+-,+-}^2(s)&=\frac{\sigma_s}{16\pi}\theta\big(s-4\mpi^2\big)\,\Sym\Big[h_{2,+-}\big(s;q_1^2,q_2^2\big)h_{2,+-}^*\big(s;q_3^2,0\big)\Big],\notag\\
 \Im \bar h_{+0,+-}^2(s)&=\frac{\sigma_s}{16\pi}\theta\big(s-4\mpi^2\big)\,\Sym\Big[h_{2,+0}\big(s;q_1^2,q_2^2\big)h_{2,+-}^*\big(s;q_3^2,0\big)\Big],\notag\\
 \Im \bar h_{0+,+-}^2(s)&=\frac{\sigma_s}{16\pi}\theta\big(s-4\mpi^2\big)\,\Sym\Big[h_{2,0+}\big(s;q_1^2,q_2^2\big)h_{2,+-}^*\big(s;q_3^2,0\big)\Big],\notag\\
 \Im \bar h_{00,+-}^2(s)&=\frac{\sigma_s}{16\pi}\theta\big(s-4\mpi^2\big)\,\Sym\Big[h_{2,00}\big(s;q_1^2,q_2^2\big)h_{2,+-}^*\big(s;q_3^2,0\big)\Big].
\end{align}
The relations~\eqref{imaginary_parts} without the bars on the left-hand side and the
$\Sym[\ldots]$ operators, defined in~\eqref{symmetrizer}, on the right-hand side simply express unitarity
for partial-wave helicity amplitudes. Since we have subtracted the FsQED
contributions and are dealing with subtracted partial-wave helicity
amplitudes, we have to correspondingly adapt the unitarity relations. This is
taken care of by the operator $\Sym[\ldots]$, which either subtracts the FsQED
contribution for charged (c) pions, or restores the symmetry factor for
neutral (n) pions
\begin{align}
\label{symmetrizer}
 \Sym\Big[h^\text{c}_{J,\lambda_1\lambda_2}\big(s;q_1^2,q_2^2\big)\Big(h^\text{c}_{J,\lambda_3\lambda_4}\big(s;q_3^2,0\big)\Big)^*\Big]&\equiv
 h^\text{c}_{J,\lambda_1\lambda_2}\big(s;q_1^2,q_2^2\big)\Big(h^\text{c}_{J,\lambda_3\lambda_4}\big(s;q_3^2,0\big)\Big)^*\notag\\
 &\quad-N_{J,\lambda_1\lambda_2}\big(s;q_1^2,q_2^2\big)N_{J,\lambda_3\lambda_4}\big(s;q_3^2,0\big),\notag\\
 \Sym\Big[h^\text{n}_{J,\lambda_1\lambda_2}\big(s;q_1^2,q_2^2\big)\Big(h^\text{n}_{J,\lambda_3\lambda_4}\big(s;q_3^2,0\big)\Big)^*\Big]&\equiv
 \frac{1}{2}h^\text{n}_{J,\lambda_1\lambda_2}\big(s;q_1^2,q_2^2\big)\Big(h^\text{n}_{J,\lambda_3\lambda_4}\big(s;q_3^2,0\big)\Big)^*.
\end{align}

The occurrence of two soft-photon zeros, both in the initial- and
final-state photon pair, implies that the convergence behavior of the
dispersion relations in~\eqref{disp_rel_diag} benefits from two
parameter-free subtractions. Since, in addition, perturbative QCD in the
factorization framework of~\cite{Brodsky_Lepage} predicts an asymptotic
behavior $\Order(1/s)$ of the helicity amplitudes for large momentum
transfer $s$, the dispersive integrals should, in principle, converge even
faster than in the case of HVP.\footnote{It should be stressed that the asymptotic behavior as predicted
by~\cite{Brodsky_Lepage} pertains to the full partial waves, including
the Born terms $N_{J,\lambda_1\lambda_2}$. However, even if
$N_{J,\lambda_1\lambda_2}$ and the Born-subtracted amplitudes
$h_{J,\lambda_1\lambda_2}-N_{J,\lambda_1\lambda_2}$ might not exhibit
the correct asymptotic behavior separately, the full amplitudes will,
provided that these constraints have been incorporated into the
calculation of $h_{J,\lambda_1\lambda_2}$ in the first place.}

At this point, 
imposing soft-photon zeros on the $\Pi_i^s$ amplitudes might appear
unjustified, since it is the full Born-subtracted helicity
amplitudes which have to obey this property, and the relation among the two
sets of quantities involves also the $\Pi_i^{t,u}$ amplitudes,
see~\eqref{eq:HelPi}. Kinematics, however, implies that if the  
direct-channel amplitudes fulfill soft-photon zeros, so do the
crossed-channel amplitudes, as we will now demonstrate. 

Consider for example the $\Pi_i^t$ amplitudes. Soft-photon constraints for
these lead to an overall prefactor $t-q_2^2$. In addition, for $q_1^2\to 0$
all $\Pi_i^t$ behave like $t-q_3^2$. Based on the $s$-channel
angle~\eqref{schannel_angle}, these factors can be rewritten as 
\begin{align}
 t-q_2^2&=\frac{s-q_3^2}{2s}\Big(q_1^2-q_2^2-s+z_s\sqrt{\lambda_{12}}\Big)\qquad\text{everywhere},\notag\\
 t-q_3^2&=-\frac{s-q_2^2}{2s}\Big(s+q_3^2-z_s\big(s-q_3^2\big)\Big)\qquad\text{for}\quad q_1^2\to 0.
\end{align}
The first equation implies not only that the $s$-channel projection of the
$t$-channel contribution has a soft-photon zero at $s=q_3^2$, but also that
the amplitude vanishes at $s=q_1^2$ for $q_2^2=0$. As the second equation
covers the opposite case, kinematics alone already ensures that soft-photon
constraints are automatically respected by the crossed-channel integrals. 

To summarize the key points: in the derivation of~\eqref{disp_rel_diag} we have
\begin{enumerate}
\item
neglected from the start any contribution with more than two pions in
intermediate states in all possible channels,
\item
separated the pion-pole and FsQED pion-loop contributions,
\item
\label{app2}
and provided a dispersive representation of the remainder in which only the
lowest partial-wave contribution to the discontinuity is kept.
\end{enumerate}
We stress that the second approximation (point~\ref{app2})
is no approximation at all if what we are
interested in is just the HLbL contribution to $a_\mu$, since contributions
from higher partial waves vanish due to angular-momentum
factors~\eqref{threshold_Dwave}. In particular, this implies that 
even in the single-meson approximation for $a_\mu$ resonances with spin larger than $2$ cannot contribute.

\subsection{Non-diagonal kernel functions}
\label{sec:non_diagonal}

While the diagonal kernel functions for the dispersion relations of the $\Pi_i$ follow immediately from $\gamma^*\gamma^*\to\pi\pi$, there can be further non-diagonal contributions, analogous to~\eqref{kernels_non_diagonal}. To derive these terms one needs to start with a set of Lorentz structures constructed in such a way that the scalar coefficient functions are free of kinematic singularities, e.g.\ for $\gamma^*\gamma^*\to\pi\pi$ these are the tensors given in~\eqref{basis}. For the $S$-waves in HLbL scattering one possible choice is
\begin{align}
 \bar \Pi^{\mu\nu\lambda\sigma}&=\tilde A_{1,s}^{\mu\nu\lambda\sigma}\tilde \Pi_{1}^s
 +\tilde A_{2,s}^{\mu\nu\lambda\sigma}\tilde \Pi_{2}^s+\text{crossed},\notag\\
 \tilde A_{1,s}^{\mu\nu\lambda\sigma}&=\big(k^\lambda q_3^\sigma-k\cdot q_3 g^{\lambda\sigma}\big)\big(q_2^\mu q_1^\nu-q_1\cdot q_2\, g^{\mu\nu}\big),\notag\\
\tilde A_{2,s}^{\mu\nu\lambda\sigma}&=\big(k^\lambda q_3^\sigma-k\cdot q_3 g^{\lambda\sigma}\big)\big(q_1\cdot q_2q_1^\mu q_2^\nu+q_1^2 q_2^2g^{\mu\nu}-q_2^2q_1^\mu q_1^\nu-q_1^2q_2^\mu q_2^\nu\big).
\end{align}
The next steps in the derivation are
\begin{enumerate}
 \item Write down unsubtracted dispersion relations for $\tilde \Pi_{i}^s$.
 \item Calculate the helicity projections of $\tilde A_{i,s}^{\mu\nu\lambda\sigma}$ and express $\tilde \Pi_{i}^s$ in terms of helicity amplitudes.
 \item Express $A_{i,s}^{\mu\nu\lambda\sigma}$ in terms of $\tilde A_{i,s}^{\mu\nu\lambda\sigma}$ and infer the dispersion relations for $\Pi_{i}^s$.
\end{enumerate}
This procedure leads to
\begin{align}
\label{disp_rel_non_diag}
 \Pi_1^s&=\frac{s-q_3^2}{\pi}\int\limits_{4\mpi^2}^\infty\frac{\diff s'}{s'-q_3^2}\bigg[\bigg(\frac{1}{s'-s}
 -\frac{s'-q_1^2-q_2^2}{\lambda_{12}'}\bigg)\Im \bar h_{++,++}^0(s')+\frac{2\xi_1\xi_2}{\lambda_{12}'} \Im \bar h^0_{00,++}(s')\bigg],\\
 -\frac{q_1^2q_2^2}{\xi_1\xi_2}\Pi_2^s&=\frac{s-q_3^2}{\pi}\int\limits_{4\mpi^2}^\infty\frac{\diff s'}{s'-q_3^2}
  \bigg[\bigg(\frac{1}{s'-s}-\frac{s'-q_1^2-q_2^2}{\lambda_{12}'}\bigg)\Im \bar h_{00,++}^0(s')
  +\frac{2q_1^2q_2^2}{\xi_1\xi_2\lambda_{12}'}\Im \bar h^0_{++,++}(s')\bigg],\notag
\end{align}
in agreement with the diagonal kernels given in~\eqref{disp_rel_diag}.
We stress that these non-diagonal kernels do not contribute to imaginary
parts in $s$: from the point of view of the analytic structure of the
amplitudes, they are polynomial contributions. 
In order to better illustrate the precise role of the
non-diagonal kernels and explain why their
presence is not tantamount to a generic subtraction polynomial, it is instructive to invert the derivation described
above. If one starts from~\eqref{disp_rel_non_diag} and inverts the
relation between the $\tilde \Pi_{i}^s$ and the helicity amplitudes, one
recovers the unsubtracted dispersion relations for the $\tilde \Pi_{i}^s$ we
started from. If one repeats the same derivation after removing the
non-diagonal kernel functions from the dispersion
relations~\eqref{disp_rel_non_diag} for the $\Pi_{i}^s$, one easily
discovers that the dispersion relations so obtained for the  $\tilde
\Pi_{i}^s$ contain kinematic singularities. We conclude that the presence
of non-diagonal kernels in the dispersion relations for $\Pi_i^s$ is
mandated by the absence of kinematic singularities in $\tilde \Pi_{i}^s$. 

The generalization of this derivation to $D$-waves requires the analog
of~\eqref{basis} for the full HLbL tensor. We derived such a basis along
the lines described in~\cite{Bardeen:1969aw,Leo:1975fb,Tarrach:1975tu}, and
will provide a full version of the dispersive system including a complete
treatment of $D$-waves in a subsequent publication~\cite{CHPS_prep}.

\subsection{Master formula}
\label{subsec:master}

The relation between the HLbL tensor and its contribution to
$a_\mu$ as given in~\eqref{lblmaster} is not valid for the $D$-wave
contributions, which involve terms ambiguous in the limit $k\to0$.  
A formula valid also in the $D$-wave case could be derived
by an expansion of the vertex function in powers of $k$, but this is impractical
since in this formulation the limit $k\to 0$ and the loop integration in general do not commute.
Therefore, we follow an approach that relies on an angular average over the spatial directions of $k$,
wherein the limit $k\to 0$ and the loop integrations may be interchanged, see App.~\ref{app:average}.\footnote{In fact, this phenomenon is an artifact of the basis we are using. It is possible to reformulate the dispersive system in such a way that no angular average is required~\cite{CHPS_prep}.} 
Performing the angular average along these lines, we obtain 
\begin{align}
\label{final_master}
 a_\mu^{\pi\pi}&=e^6\int\frac{\diff^4q_1}{(2\pi)^4}\int\frac{\diff^4q_2}{(2\pi)^4}\frac{I^{\pi
   \pi}}{q_1^2q_2^2s\big((p+q_1)^2-m^2\big)
 \big((p-q_2)^2-m^2\big)}, \notag \\
I^{\pi \pi}&= \sum\limits_{i\in\{1,2,3,6,14\}}\Big(T_{i,s} I_{i,s}+2T_{i,u} I_{i,u}\Big) + 2T_{9,s}I_{9,s}+2T_{9,u}I_{9,u}+2T_{12,u}I_{12,u},
\end{align}
with dispersive integrals $I_{i,(s,u)}$ given in App.~\ref{app:master_int} and
integration kernels $T_{i,(s,u)}$ in App.~\ref{app:master_kernels}. 
Throughout, we used the symmetry of the integrand under $q_1\leftrightarrow
-q_2$ to map the $t$-channel contributions onto the $u$-channel and
simplify the $s$-channel kernels. Moreover, this symmetry transforms the
amplitudes corresponding to $h^2_{+0,+-}$ and $h^2_{0+,+-}$ into each
other, with the $t$-channel of one equaling the $u$-channel of the other,
and makes the $s$-channel contribution of $h^2_{0+,+-}$ coincide with the
one generated by $h^2_{+0,+-}$. 

\begin{figure}
\centering
 \includegraphics[width=2.5cm,clip]{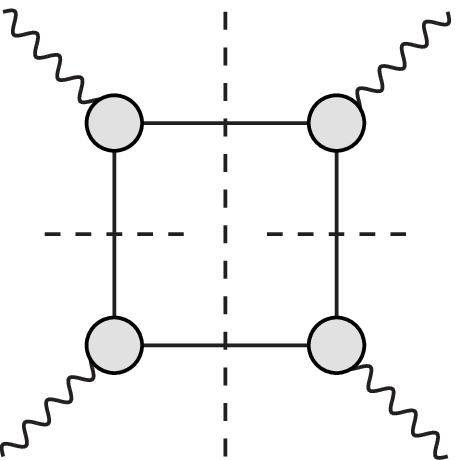}\quad
 \includegraphics[width=2.5cm,clip]{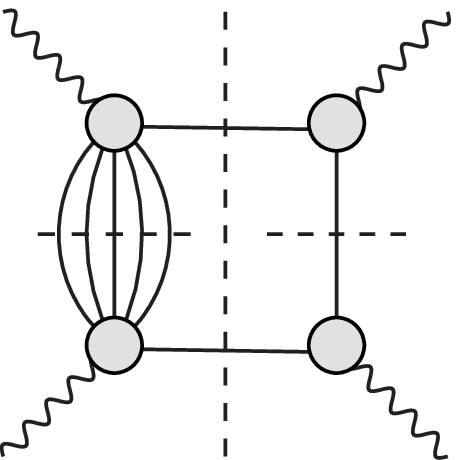}\quad
 \includegraphics[width=2.5cm,clip]{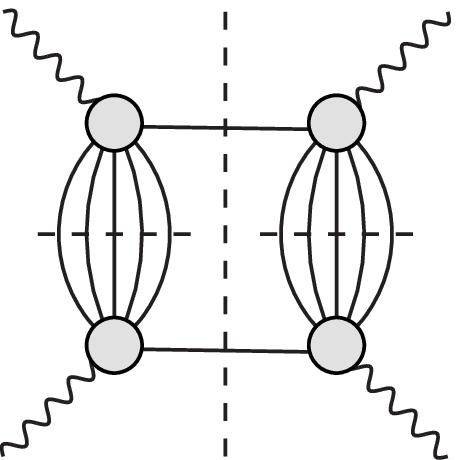}\quad
 \includegraphics[width=2.5cm,clip]{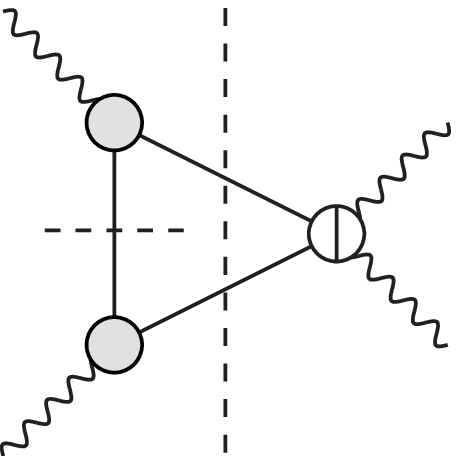}\quad
 \includegraphics[width=2.5cm,clip]{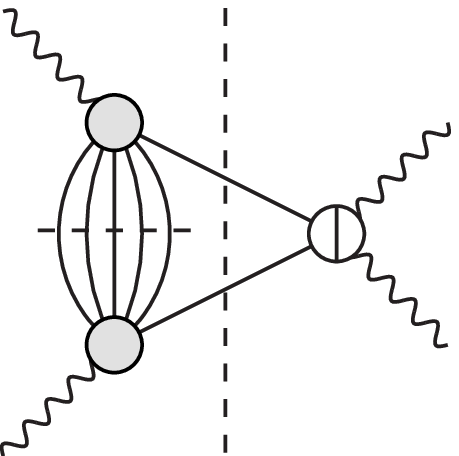}\quad
 \raisebox{0.57cm}{\includegraphics[width=2.5cm,clip]{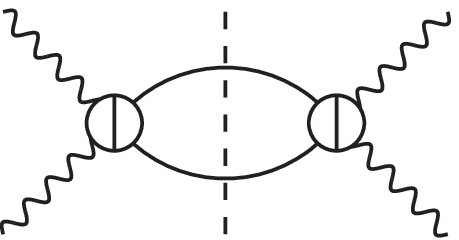}}
\caption{Classes of unitarity diagrams in HLbL scattering. The grey blobs
  denote (transition) form factors and the blobs with vertical line a
  polynomial contribution in the crossed channel.} 
\label{fig:unitarity}
\end{figure}

The use and interpretation of~\eqref{final_master} requires some
discussion. In particular, we return to the separation of the
FsQED term from the rest which we introduced in
Sect.~\ref{sec:disprel}. To implement this separation correctly and avoid
double counting, we must specify what we mean by the partial waves of
$\gamma^*\gamma^*\to\pi^+\pi^-$ which appear in~\eqref{final_master}
via the $I_{i,(s,u)}$ (see App.~\ref{app:master_int} and~\eqref{imaginary_parts}). 
We decompose the charged-pion partial waves
into Born terms $N_{J,\lambda_1\lambda_2}$ (see App.~\ref{app:FsQED}) and a remainder $\hat
h_{J,\lambda_1\lambda_2}$ to obtain the decomposition
($i=\lambda_1\lambda_2$, $j=\lambda_3\lambda_4$)
\begin{align}
\label{Born_separate}
 h_{J,i}\big(s;q_1^2,q_2^2\big)h_{J,j}^*\big(s;q_3^2,0\big)&= 
  N_{J,i}\big(s;q_1^2,q_2^2\big)N_{J,j}\big(s;q_3^2,0\big)+
   N_{J,i}\big(s;q_1^2,q_2^2\big)\hat h_{J,j}^*\big(s;q_3^2,0\big)\notag\\
   &+
  \hat h_{J,i}\big(s;q_1^2,q_2^2\big)N_{J,j}\big(s;q_3^2,0\big)+
   \hat h_{J,i}\big(s;q_1^2,q_2^2\big)\hat h_{J,j}^*\big(s;q_3^2,0\big).
\end{align}
The first term has to be removed from all the $I_{i,(s,u)}$  entering~\eqref{final_master} 
since it is accounted for by the FsQED
charged pion loop, see~\eqref{symmetrizer}. The second and third term correspond to a triangle-type
contribution (fourth diagram in Fig.~\ref{fig:unitarity}), and the last
term to a bulb topology (last diagram in
Fig.~\ref{fig:unitarity}).\footnote{In sQED the occurrence of the seagull
  term mandated by gauge invariance implies that $N_{J,i}$ includes certain
  non-pole pieces, which gives rise to triangle and bulb
  topologies. However, to ensure gauge invariance at each step, these
  contributions should be kept within the FsQED part of the calculation.}
Physically, the amplitudes $\hat h_{J,\lambda_1\lambda_2}$
include for instance $\pi\pi$ rescattering effects, and thus allow for a model-independent
treatment of degrees of freedom  corresponding to resonances coupling to the $\pi\pi$ channel, such as the $\sigma$-resonance.

We now discuss in more detail the precise meaning of our initial statement
that in our dispersive approach we neglected multi-pion
intermediate states. Consider the different classes of unitarity
diagrams that involve $\pi\pi$ intermediate states in the $s$-channel, as
shown in Fig.~\ref{fig:unitarity}. Although, by definition, all these
diagrams are $\pi\pi$ reducible in the $s$-channel, they differ in the
analytic structure in the crossed channel: sub-diagrams feature a pion pole,
multi-pion exchange, or polynomial contributions. 
Our dispersive representation of the HLbL amplitude was derived in a
partial-wave picture and therefore cannot, by construction, have any
crossed-channel cut.
Box diagrams with multi-pion exchange involve crossed-channel multi-pion
cuts, representing intermediate states with more than two pions. 
These unitarity diagrams are not neglected completely, but only included in
a partial-wave approximation. In this framework, the second diagram in
Fig.~\ref{fig:unitarity} is partially covered by the second and third term
in~\eqref{Born_separate}, because the partial-wave projection of the
left-hand side of the diagram (which contains the multi-pion exchange) is
contained in $\hat h_{J,j}$. Analogously the third and fifth diagram belong to the
last term therein. Indeed, if the first term 
in~\eqref{Born_separate} is kept, the contribution from the dispersive integrals 
in~\eqref{final_master} 
exactly corresponds to the first term of the
partial-wave expansion  of the charged-pion loop calculated within sQED
multiplied by the appropriate power of the pion vector form factor.

The subtleties concerning Born terms are absent for neutral pions. However,
since we have assumed distinguishable particles in the derivation of the
unitarity relation, their contribution has to be accompanied by a symmetry
factor of $1/2$, see~\eqref{symmetrizer}, so that the explicit form of the
$\gamma^*\gamma^*\to\pi\pi$ amplitudes occurring in the imaginary
part reads 
\begin{align}
\label{pipi_states}
&\Sym\Big[h_{J,i}^\text{c}\big(s;q_1^2,q_2^2\big)\Big(h_{J,j}^\text{c}\big(s;q_3^2,0\big)\Big)^*\Big]
+\Sym\Big[h_{J,i}^\text{n}\big(s;q_1^2,q_2^2\big)\Big(h_{J,j}^\text{n}\big(s;q_3^2,0\big)\Big)^*\Big]\\
&=
h_{J,i}^\text{c}\big(s;q_1^2,q_2^2\big)\Big(h_{J,j}^\text{c}\big(s;q_3^2,0\big)\Big)^*+
\frac{1}{2}h_{J,i}^\text{n}\big(s;q_1^2,q_2^2\big)\Big(h_{J,j}^\text{n}\big(s;q_3^2,0\big)\Big)^* 
-N_{J,i}\big(s;q_1^2,q_2^2\big)N_{J,j}\big(s;q_3^2,0\big),\notag
\end{align}
adding the contributions from charged and neutral pions and
subtracting the FsQED part. Alternatively, \eqref{pipi_states} may be
expressed in the isospin basis. Changing basis towards isospin $0$ and $2$
according to~\cite{HPS} 
\beq
\begin{pmatrix}
h_{J,i}^\text{c}\\ h_{J,i}^\text{n} 
\end{pmatrix}
=\begin{pmatrix}
  \frac{1}{\sqrt{3}} & \frac{1}{\sqrt{6}}\\
  \frac{1}{\sqrt{3}} & -\sqrt{\frac{2}{3}}
 \end{pmatrix}
\begin{pmatrix}
h_{J,i}^0\\ h_{J,i}^2 
\end{pmatrix},
\eeq
we find
\begin{align}
\label{ima_isospin_basis}
&\Sym\Big[h_{J,i}^\text{c}\big(s;q_1^2,q_2^2\big)\Big(h_{J,j}^\text{c}\big(s;q_3^2,0\big)\Big)^*\Big]
+\Sym\Big[h_{J,i}^\text{n}\big(s;q_1^2,q_2^2\big)\Big(h_{J,j}^\text{n}\big(s;q_3^2,0\big)\Big)^*\Big]\\
&=
\frac{1}{2}h_{J,i}^{0}\big(s;q_1^2,q_2^2\big)\Big(h_{J,j}^{0}\big(s;q_3^2,0\big)\Big)^*+
\frac{1}{2}h_{J,i}^{2}\big(s;q_1^2,q_2^2\big)\Big(h_{J,j}^{2}\big(s;q_3^2,0\big)\Big)^*
-N_{J,i}\big(s;q_1^2,q_2^2\big)N_{J,j}\big(s;q_3^2,0\big).\notag
\end{align}
As long as the virtualities remain below the two-pion threshold, Watson's
theorem~\cite{Watson} implies that the phase of
$h_{J,i}^I\big(s;q_1^2,q_2^2\big)$ is equal to $\delta_J^I(s)$, the phase
shift of the corresponding $\pi\pi$ partial wave. Since the Born terms
$N_{J,i}\big(s;q_1^2,q_2^2\big)$ are real, this proves that indeed the
final result for the imaginary part is a real quantity. 

\begin{figure}
\centering
 \includegraphics[width=4cm,clip]{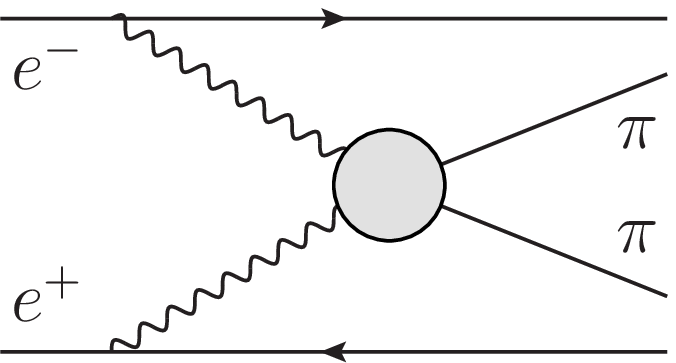}\qquad\qquad
 \includegraphics[width=4cm,clip]{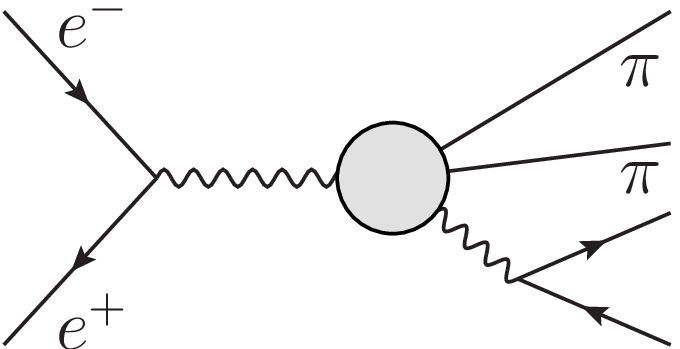}
\caption{$\gamma^*\gamma^*\to\pi\pi$ physics in space-like (left) and
  time-like (right) kinematics.} 
\label{fig:eepipi}
\end{figure}

Experimentally, $\gamma^*\gamma^*\to\pi\pi$ partial waves are accessible in
the process $e^+e^-\to e^+e^-\pi\pi$, either in space-like or time-like
configuration, see Fig.~\ref{fig:eepipi}. Using similar manipulations of
the loop integrals as in the case of the pion pole~\cite{KN,JN},  
only negative virtualities are required for HLbL scattering. However,
within the dispersive approach information from all kinematic regions is
highly welcome to provide additional experimental constraints and
potentially improve the accuracy, as dispersion theory is the ideal
framework to reliably perform the required analytic continuation. In this
particular case, the construction of dispersion relations for the
doubly-virtual process $\gamma^*\gamma^*\to\pi\pi$ in time-like kinematics
is complicated by the occurrence of anomalous
thresholds~\cite{Mandelstam,LMS}. For a description of how to account for
those effects within our framework see~\cite{HCPS}.  

In summary, we obtain a decomposition of the full HLbL contribution into
\beq
a_\mu^\text{HLbL}=a_\mu^{\pi^0\text{-pole}} + a_\mu^\text{FsQED} + a_\mu^{\pi\pi},
\label{amu_total}
\eeq
with the pion pole~\eqref{amu_pole}, the FsQED charged pion 
loop,\footnote{An explicit representation is provided in
  App.~\ref{app:FsQED}. For the derivation of the sQED contribution we
  refer to~\cite{Kinoshita:1984it,Hayakawa:1996ki,Kuhn:2003pu} and for a
  calculation of higher chiral orders to~\cite{Engel:2012xb}.} and the
remaining effects from $\pi\pi$ intermediate states according
to~\eqref{final_master} with imaginary parts as given
in~\eqref{pipi_states} or~\eqref{ima_isospin_basis} in particle or isospin
basis, respectively. $a_\mu^{\pi\pi}$ covers unitarity diagrams both with 
triangle and bulb topologies. As far as FsQED Born terms are concerned, this
is exemplified by the decomposition in~\eqref{Born_separate}. It should be
stressed that both the pion loop and the Born-term contribution to the
triangle topologies in $a_\mu^{\pi\pi}$ are entirely fixed by the pion
vector form factor for the $\pi^+\pi^-\gamma^*$ vertex. Due to the fact
that only space-like kinematics are relevant for $a_\mu$, the dispersive
integrals also for triangle topologies are free of anomalous
thresholds. But even if they were not, the present formalism would still be
useful: one would merely have to extend the framework along the lines 
sketched in~\cite{HCPS}.

\section{Conclusions and outlook}
\label{sec:outlook}

The main result of this paper is presented in~\eqref{amu_total} and
\eqref{final_master}: a master formula that gives the contribution from
$\pi\pi$ intermediate states in HLbL scattering to the anomalous magnetic
moment of the muon, expressed in terms of helicity partial waves for
$\gamma^*\gamma^*\to\pi\pi$. Within the general framework of dispersion
theory, $\pi\pi$ intermediate states constitute the second most important
contribution after pseudoscalar pole terms, also included
in~\eqref{amu_total}. This result is important progress towards
a model-independent, data-driven analysis of HLbL scattering. A first
numerical evaluation of these contributions as they appear
in~\eqref{amu_total} and~\eqref{final_master} as well as a generalization to 
a complete treatment of $D$-waves is in
progress~\cite{CHPS_prep}.

Although in principle the required helicity partial waves for
$\gamma^*\gamma^*\to\pi\pi$ can be measured, a dispersive treatment of
$\gamma^*\gamma^*\to\pi\pi$ in the framework of Roy--Steiner equations
allows for a combined analysis of all experimental constraints available
for the relevant pion--photon interactions~\cite{HPS,CHPS_prep}, in
particular from kinematic regions different from those directly relevant for
 HLbL scattering. While heavier two-particle, scalar
intermediate states (such as $K\bar K$) are amenable to the same treatment, it is 
more challenging to account for multi-pion contributions at the same
level of rigor that we have adhered to here. To estimate the impact of
multi-pion intermediate states, possible ans\"atze would be to try to
generalize the calculation of the FsQED pion loop to include resonances or
to approximate missing physical degrees of freedom in terms of resonance
poles along the lines of~\cite{Pauk}.

The final goal of the approach laid out here is a calculation of HLbL
scattering consistent with the general principles of analyticity,
unitarity, crossing symmetry, and gauge invariance and backed by data as
closely as possible. To this end, also the pseudoscalar transition form
factors should be subject to a similar analysis,
see~\cite{Czerwinski,SKN,MesonNet,Hanhart_eta} for first steps in this
direction. Ultimately, this approach should allow for a more reliable
estimate of uncertainties in the HLbL contribution to the anomalous
magnetic moment of the muon.

\subsection*{Acknowledgements}

We would like to thank Andreas Nyffeler for helpful discussions, and Bastian Kubis, Heiri Leutwyler, and Andreas Nyffeler
for comments on the manuscript.
Financial support by the Swiss National Science Foundation is gratefully acknowledged.
The AEC is supported by the
  ``Innovations- und Kooperationsprojekt C-13'' of the ``Schweizerische
  Universit\"atskonferenz SUK/CRUS.''
  
\appendix

\section{FsQED Born terms and contribution to $\boldsymbol{a_\mu}$}
\label{app:FsQED}

In our conventions, the $S$- and $D$-wave projections of the Born terms for $\gamma^*\gamma^*\to\pi\pi$ read~\cite{CHPS_prep}
\begin{align}
 N_{J,++}\big(t;q_1^2,q_2^2\big)&=\Bigg\{\frac{2}{p_tq_t}Q_J(x_t)\bigg(\mpi^2+\frac{tq_1^2q_2^2}{\lambda_{12}^t}\bigg)+2\delta_{J0}\frac{\big(q_1^2-q_2^2\big)^2-t\big(q_1^2+q_2^2\big)}{\lambda_{12}^t}\Bigg\}F_\pi^V\big(q_1^2\big)F_\pi^V\big(q_2^2\big),\notag\\
N_{2,+-}\big(t;q_1^2,q_2^2\big)&=\Bigg\{-\frac{2}{p_tq_t}\bigg(d_{20}^2(x_t)Q_0(x_t)+\frac{\sqrt{6}}{4}x_t\bigg)
\bigg(\mpi^2+\frac{tq_1^2q_2^2}{\lambda_{12}^t}\bigg)
+\frac{2}{\sqrt{6}}\frac{t\big(t-q_1^2-q_2^2\big)}{\lambda_{12}^t}\Bigg\}\notag\\
&\qquad\times F_\pi^V\big(q_1^2\big)F_\pi^V\big(q_2^2\big),\notag\\
N_{2,+0}\big(t;q_1^2,q_2^2\big)&=-\frac{t+q_1^2-q_2^2}{t-q_1^2-q_2^2}\frac{q_2^2}{\xi_2}\sqrt{\frac{3}{t}}\frac{2p_t}{q_t}\bigg\{x_t^2\big(1-x_t^2\big)Q_0(x_t)+x_t\bigg(x_t^2-\frac{2}{3}\bigg)\bigg\}F_\pi^V\big(q_1^2\big)F_\pi^V\big(q_2^2\big),\notag\\
N_{2,0+}\big(t;q_1^2,q_2^2\big)&=-\frac{t-q_1^2+q_2^2}{t-q_1^2-q_2^2}\frac{q_1^2}{\xi_1}\sqrt{\frac{3}{t}}\frac{2p_t}{q_t}\bigg\{x_t^2\big(1-x_t^2\big)Q_0(x_t)+x_t\bigg(x_t^2-\frac{2}{3}\bigg)\bigg\}F_\pi^V\big(q_1^2\big)F_\pi^V\big(q_2^2\big),\notag\\
N_{J,00}\big(t;q_1^2,q_2^2\big)&=-\frac{q_1^2q_2^2}{\xi_1\xi_2\lambda_{12}^t}\bigg\{\Big(t^2-\big(q_1^2-q_2^2\big)^2\Big)\frac{1}{p_tq_t}Q_J(x_t)-8t\delta_{J0}\bigg\}F_\pi^V\big(q_1^2\big)F_\pi^V\big(q_2^2\big),
\end{align}
with
\beq
x_t=\frac{t-q_1^2-q_2^2}{4p_tq_t},
\eeq
and the lowest Legendre functions of the second kind
\beq
Q_0(z)=\frac{1}{2}\int\limits_{-1}^1\frac{\diff x}{z-x},\qquad Q_0(z\pm i\eps)=\frac{1}{2}\log\bigg|\frac{1+z}{1-z}\bigg|\mp i\frac{\pi}{2}\theta\big(1-z^2\big),\qquad Q_2(z)=P_2(z)Q_0(z)-\frac{3}{2}z.
\eeq
$F_\pi^V(q^2)$ denotes the pion vector form factor and the kinematic quantities are defined as in Sect.~\ref{sec:ggpipi}.

However, for the reasons discussed in Sects.~\ref{sec:disprel} and~\ref{subsec:master} the FsQED contribution to $a_\mu$ cannot be analyzed within a partial-wave framework, but rather based on Feynman integrals. 
We find the representation
\beq
a_\mu^\text{FsQED}=\frac{2e^6}{3\pi^2}\int\frac{\diff^4q_1}{(2\pi)^4}\int\frac{\diff^4q_2}{(2\pi)^4}
\frac{F_\pi^V\big(q_1^2\big)F_\pi^V\big(q_2^2\big)F_\pi^V(s)\big(I_s+2I_u+J_1+J_2\big)}{q_1^2q_2^2s\big((p+q_1)^2-m^2\big)\big((p-q_2)^2-m^2\big)},
\eeq
where
\begin{align}
 I_s&=\Big\{m^2 s-2p\cdot q_1\big(p\cdot q_1+p\cdot q_2-q_1\cdot q_2-q_2^2\big)\Big\}\int\limits_0^1\diff x\frac{x(1-x)}{\mpi^2-x(1-x)s},\\
 I_u&=\Big\{q_1^2\big(m^2+p\cdot q_2\big)-p\cdot q_1\big(p\cdot q_1+q_1\cdot q_2\big)\Big\}\int\limits_0^1\diff x\frac{x(1-x)}{\mpi^2-x(1-x)q_1^2},\notag\\
 J_1&=\int\limits_0^1\diff x\int\limits_0^{1-x}\diff y\,\Delta^{-1}(x,y)
 \Bigg\{2(16x y-6x-6y+3)p\cdot q_1\, p\cdot q_2-4x(8x-5)(p\cdot q_1)^2\notag\\
 &+2p\cdot q_1\Big[2x(5x+2y-3)q_1\cdot q_2-\big(8x y+2y^2-6x-6y+3\big)q_2^2\Big]
 -8m^2x(y\, q_1\cdot q_2+(1-x)q_1^2\big)\Bigg\},\notag\\
 J_2&=\int\limits_0^1\diff x\int\limits_0^{1-x}\diff y\,\Delta^{-2}(x,y)
 \Bigg\{\frac{m^2}{2}xy(1-x-y)\lambda_{12}\notag\\
 &+2(p\cdot q_1)^2\Big[x(x+2y-1)q_1\cdot q_2+y\big(2x(x+y-1)+1-y\big)q_2^2\Big]\notag\\
 &-p\cdot q_1\,p\cdot q_2\Big[\big(x^2(4y-1)+x(1-2y)^2+y(1-y)\big)q_1\cdot q_2+x(x+2y-1)q_1^2+y(2x+y-1)q_2^2\Big]\notag\\
 &+p\cdot q_1\bigg[y\Big[(1-2y)^2(2x+y-1)q_2^2-(2x-1)\big(4x^2+6x(y-1)+1-y\big)q_1^2\Big]q_2^2\notag\\
 &+x\Big[8(x-1)x y\, q_1\cdot q_2+(1-2x)^2(x+2y-1)q_1^2-\big(4(3x-2)y^2+(8y-1)(1-x)\big)q_2^2\Big]q_1\cdot q_2\bigg]\Bigg\},\notag
\end{align}
and
\beq
\Delta(x,y)=\mpi^2-x y s-x(1-x-y)q_1^2-y(1-x-y)q_2^2.
\eeq

\section{$\boldsymbol{1}$-loop ChPT for $\boldsymbol{\gamma^*\gamma^*\to\pi\pi}$}
\label{app:ggpipi_ChPT}

At $1$-loop order the only non-vanishing partial-wave amplitudes are (for the on-shell case see~\cite{Bijnens:1987dc,Donoghue:1988eea})
\begin{align}
\label{ggpipi_ChPT}
h_{0,++}\big(t;q_1^2,q_2^2\big)&=\frac{\bar l_6-\bar l_5}{48\pi^2\Fpi^2}\big(t-q_1^2-q_2^2\big)\begin{Bmatrix}
                                      1\\ 0
                                     \end{Bmatrix}
-\frac{1}{8\pi^2\Fpi^2}\begin{Bmatrix}
                                      t/2\\ t-\mpi^2
                                     \end{Bmatrix}
\Bigg\{1+2\bigg(\mpi^2+\frac{tq_1^2q_2^2}{\lambda_{12}^t}\bigg)C_0\big(t,q_1^2,q_2^2\big)\notag\\
&+\frac{t\big(q_1^2+q_2^2\big)-\big(q_1^2-q_2^2\big)^2}{\lambda_{12}^t}\bar J(t)
-\frac{q_1^2\big(t+q_2^2-q_1^2\big)}{\lambda_{12}^t}\bar J\big(q_1^2\big)
-\frac{q_2^2\big(t+q_1^2-q_2^2\big)}{\lambda_{12}^t}\bar J\big(q_2^2\big)
\Bigg\},\notag\\
h_{0,00}\big(t;q_1^2,q_2^2\big)&=\frac{q_1^2q_2^2}{\xi_1\xi_2}\Bigg[\frac{\bar l_6-\bar l_5}{24\pi^2\Fpi^2}\begin{Bmatrix}
                                      1\\ 0
                                     \end{Bmatrix}
+\frac{1}{8\pi^2\Fpi^2\lambda_{12}^t}\begin{Bmatrix}
                                      t/2\\ t-\mpi^2
                                     \end{Bmatrix}
\bigg\{\Big(t^2-\big(q_1^2-q_2^2\big)^2\Big)C_0\big(t,q_1^2,q_2^2\big)\notag\\
&+4t\bar J(t)-2\big(t+q_1^2-q_2^2\big)\bar J\big(q_1^2\big)-2\big(t-q_1^2+q_2^2\big)\bar J\big(q_2^2\big)\bigg\}\Bigg],                                     
\end{align}
where the upper/lower result refers to charged/neutral pions and the loop functions are defined as
\begin{align}
 C_0\big(t,q_1^2,q_2^2\big)&=-\int\limits_0^1\diff x\int\limits_0^{1-x}\diff y\frac{1}{\mpi^2-t x y-q_1^2 x(1-x-y)-q_2^2y(1-x-y)},\notag\\
 \bar J(t)&=-\int\limits_0^1\diff x\log\bigg[1-x(1-x)\frac{t}{\mpi^2}\bigg],
\end{align}
with imaginary parts
\beq
\Imt \bar J(t)=\pi \sigma_t\theta\big(t-4\mpi^2\big),\qquad \Imt C_0\big(t,q_1^2,q_2^2\big)=-\frac{\pi \theta\big(t-4\mpi^2\big)}{\sqrt{\lambda_{12}^t}}
\log\frac{t-q_1^2-q_2^2+\sigma_t\sqrt{\lambda_{12}^t}}{t-q_1^2-q_2^2-\sigma_t\sqrt{\lambda_{12}^t}}.
\eeq
In the limit $q_i^2\to 0$, \eqref{ggpipi_ChPT} reproduces the leading term of the chiral expansion of the pion polarizabilities
\beq
 h_{0,++}(t;0,0)=\frac{\mpi}{2\alpha}\big(\alpha_1-\beta_1\big)t+\Order\big(t^2\big),\qquad
 \frac{\mpi}{2\alpha}\big(\alpha_1-\beta_1\big)=\frac{1}{96\pi^2\Fpi^2}\begin{Bmatrix}
                                      2(\bar l_6-\bar l_5)\\ -1
                                     \end{Bmatrix}.
\eeq
In practice, subtractions have to be introduced into the Roy--Steiner equations for $\gamma^*\gamma^*\to\pi\pi$ sketched in Sect.~\ref{sec:ggpipi} in order to suppress the high-energy tail of the integrals. The ChPT result~\eqref{ggpipi_ChPT} as well as
experimental information on pion polarizabilities provide valuable constraints for this subtraction term.
Moreover, the $1$-loop chiral amplitudes themselves can be used to illustrate the $S$-wave part of the Roy--Steiner system~\eqref{kernels_diagonal} and~\eqref{kernels_non_diagonal}. Inserting~\eqref{ggpipi_ChPT} into these equations, one finds the relations\footnote{Due to the high-energy behavior of the ChPT amplitudes a subtraction is required. The subsequent relations follow e.g.\ from the charged channel when subtracting at $t=0$. Note that the dispersive integrals in the formulation given here apply to the situation where the virtualities are sufficiently small that no anomalous thresholds occur. If anomalous thresholds are present, the integration contour has to be deformed as described in~\cite{HCPS}.} 
\begin{align}
\label{kernel_relations}
&\frac{1}{\pi}\int\limits_{4\mpi^2}^\infty\diff t'\Bigg\{\bigg(\frac{1}{t'-t}-\frac{t'-q_1^2-q_2^2}{\lambda_{12}^{t'}}\bigg)
\Imt h_1\big(t';q_1^2,q_2^2\big)
 +\frac{2q_1^2q_2^2}{\lambda_{12}^{t'}}\Imt h_2\big(t';q_1^2,q_2^2\big)\Bigg\}\\
 &=1+2\bigg(\mpi^2+\frac{tq_1^2q_2^2}{\lambda_{12}^t}\bigg)C_0\big(t,q_1^2,q_2^2\big)
 +\frac{t\big(q_1^2+q_2^2\big)-\big(q_1^2-q_2^2\big)^2}{\lambda_{12}^t}\bar J(t)\notag\\
 &\quad-\frac{q_1^2\big(t+q_2^2-q_1^2\big)}{\lambda_{12}^t}\bar J\big(q_1^2\big)
-\frac{q_2^2\big(t+q_1^2-q_2^2\big)}{\lambda_{12}^t}\bar J\big(q_2^2\big),\notag\\
&\frac{1}{\pi}\int\limits_{4\mpi^2}^\infty\diff t'\Bigg\{\bigg(\frac{1}{t'-t}-\frac{t'-q_1^2-q_2^2}{\lambda_{12}^{t'}}\bigg)
\Imt h_2\big(t';q_1^2,q_2^2\big)
 +\frac{2}{\lambda_{12}^{t'}}\Imt h_1\big(t';q_1^2,q_2^2\big)\Bigg\}\notag\\
 &=-\frac{1}{\lambda_{12}^t}\bigg\{\Big(t^2-\big(q_1^2-q_2^2\big)^2\Big)C_0\big(t,q_1^2,q_2^2\big)
 +4t\bar J(t)-2\big(t+q_1^2-q_2^2\big)\bar J\big(q_1^2\big)-2\big(t-q_1^2+q_2^2\big)\bar J\big(q_2^2\big)\bigg\},\notag
 \end{align}
 where $h_1$ and $h_2$ with imaginary parts
 \begin{align}
 \Imt h_1\big(t;q_1^2,q_2^2\big)&=2\bigg(\mpi^2+\frac{tq_1^2q_2^2}{\lambda_{12}^{t}}\bigg)\Imt C_0\big(t,q_1^2,q_2^2\big)
+\frac{t\big(q_1^2+q_2^2\big)-\big(q_1^2-q_2^2\big)^2}{\lambda_{12}^{t}}\Imt \bar J(t),\notag\\
\Imt h_2\big(t;q_1^2,q_2^2\big)&=-\frac{1}{\lambda_{12}^t}\Big[\Big(t^2-\big(q_1^2-q_2^2\big)^2\Big)\Imt C_0\big(t,q_1^2,q_2^2\big)+4t\Imt \bar J(t)\Big],
\end{align}
correspond to $h_{0,++}$ and $h_{0,00}$, respectively.
We checked numerically that the relations~\eqref{kernel_relations} hold. In particular, we find that in general the contribution from the non-diagonal kernels is non-negligible.

\section{Lorentz structures for the HLbL tensor}
\label{app:Lorentz}

The Lorentz structures that appear in the unitarity relation~\eqref{Imsresult} are
\begin{align}
A_{1,s}^{\mu\nu\lambda\sigma}&=\frac{8}{\big(s-q_3^2\big)\lam}\Big(k^\lambda q_3^\sigma-k\cdot q_3\, g^{\lambda\sigma}\Big)\bigg(q^{\mu\nu}_{12}+\frac{\lam}{4}g^{\mu\nu}\bigg),\notag\\
A_{2,s}^{\mu\nu\lambda\sigma}&=\frac{8}{\big(s-q_3^2\big)\lam}\Big(k^\lambda q_3^\sigma-k\cdot q_3\, g^{\lambda\sigma}\Big)
\Big(q_1\cdot q_2\,q_1^\mu-q_1^2\,q_2^\mu\Big)\Big(q_2^2\,q_1^\nu-q_1\cdot q_2\,q_2^\nu\Big),\notag\\
A_{3,s}^{\mu\nu\lambda\sigma}&=-\frac{8}{5\big(s-q_3^2\big)\lam}\bigg\{\big(1-z_s^2\big)\Big(k^\lambda q_3^\sigma-k\cdot q_3\,g^{\lambda\sigma}\Big)
+\frac{4s\,q_{tu}^\lambda q_{tu}^\sigma}{\big(s-q_3^2\big)\lam}\bigg\}\bigg(q^{\mu\nu}_{12}+\frac{\lam}{4}g^{\mu\nu}\bigg),\notag\\
A_{4,s}^{\mu\nu\lambda\sigma}&=\frac{32\bar z_s\,q_{tu}^\sigma}{5\big(s-q_3^2\big)^2\lam}\Big(k\cdot q_3\,q_3^\lambda-q_3^2\,k^\lambda\Big)
\bigg(q^{\mu\nu}_{12}+\frac{\lam}{4}g^{\mu\nu}\bigg),\notag\\
A_{5,s}^{\mu\nu\lambda\sigma}&=-\frac{4}{5\big(s-q_3^2\big)}\Big(k^\lambda q_3^\sigma-k\cdot q_3\,g^{\lambda\sigma}\Big)
\Bigg\{\frac{1}{2}\big(1-z_s^2\big)g^{\mu\nu}-2\bar z_s\big(q_1^\mu q_1^\nu-q_2^\mu q_2^\nu\big)
\notag\\
&+\big(q_1^\mu+q_2^\mu\big)\big(q_1^\nu+q_2^\nu\big)\frac{1}{s}\Big(1+z_s^2+2\bar z_s\big(q_1^2-q_2^2\big)\Big)
+\frac{2\big(k^\mu k_s^\nu+k_s^\mu k^\nu\big)}{s-q_3^2}
+\frac{2}{\lam}\big(1+z_s^2\big)q^{\mu\nu}_{12}\Bigg\},\notag\\
A_{6,s}^{\mu\nu\lambda\sigma}&=\frac{16}{15\big(s-q_3^2\big)^2\lam}\Bigg\{-\frac{\big(s-q_3^2\big)^2\lam}{4}\big(g^{\mu\lambda}g^{\nu\sigma}+g^{\mu\sigma}g^{\nu\lambda}\big)\notag\\
&+\frac{s-q_3^2}{2}\Bigg[g^{\mu\nu}\bigg[\frac{\lam}{2}\big(z_s^2-3\big)\Big(k^\lambda q_3^\sigma-k\cdot q_3\,g^{\lambda\sigma}\Big)-\frac{2s}{s-q_3^2}\,q_{tu}^\lambda q_{tu}^\sigma\bigg]\notag\\
&+g^{\mu\lambda}\bigg[\lam\Big(k^\nu q_3^\sigma-k\cdot q_3\,g^{\nu\sigma}\Big)-q_{tu}^\sigma\Big(\big(s-q_1^2+q_2^2\big)q_1^\nu-\big(s+q_1^2-q_2^2\big)q_2^\nu\Big)\bigg]\notag\\
&+g^{\nu\lambda}\bigg[\lam\Big(k^\mu q_3^\sigma-k\cdot q_3\,g^{\mu\sigma}\Big)-q_{tu}^\sigma\Big(\big(s-q_1^2+q_2^2\big)q_1^\mu-\big(s+q_1^2-q_2^2\big)q_2^\mu\Big)\bigg]\notag\\
&\hspace{-27pt}+g^{\mu\sigma}\bigg[\lam\Big(q_3^\nu k^\lambda-k\cdot q_3\,g^{\nu\lambda}\Big)+q_{tu}^\lambda\Big(\big(q_1^2-q_2^2\big)\big(q_1^\nu+q_2^\nu\big)-s\big(q_1^\nu-q_2^\nu\big)\Big)-\frac{2\lam\,k^\nu}{s-q_3^2}\Big(k\cdot q_3\,q_3^\lambda-q_3^2\,k^\lambda\Big)\bigg]\notag\\
&\hspace{-27pt}+g^{\nu\sigma}\bigg[\lam\Big(q_3^\mu k^\lambda-k\cdot q_3\,g^{\mu\lambda}\Big)+q_{tu}^\lambda\Big(\big(q_1^2-q_2^2\big)\big(q_1^\mu+q_2^\mu\big)-s\big(q_1^\mu-q_2^\mu\big)\Big)-\frac{2\lam\,k^\mu}{s-q_3^2}\Big(k\cdot q_3\,q_3^\lambda-q_3^2\,k^\lambda\Big)\bigg]\Bigg]\notag\\
&-\lam \Big(k^\lambda q_3^\sigma-k\cdot q_3\,g^{\lambda\sigma}\Big)\Bigg[k^\mu k_s^\nu+k_s^\mu k^\nu
+\frac{s-q_3^2}{2s}\bigg[\Big(1+z_s^2+2\bar z_s\big(q_1^2-q_2^2\big)\Big)\big(q_1^\mu+q_2^\mu\big)\big(q_1^\nu+q_2^\nu\big)\notag\\
&\qquad-2s\bar z_s\big(q_1^\mu q_1^\nu-q_2^\mu q_2^\nu\big)+\frac{2s}{\lam}\big(z_s^2+3\big)q_{12}^{\mu\nu}\bigg]\Bigg]\notag\\
&\hspace{-8pt}+2q_{tu}^\sigma \bigg[q_{tu}^\lambda\Big(\big(q_1^\mu+q_2^\mu\big)\big(q_1^\nu+q_2^\nu\big)+\frac{2s}{\lam}q_{12}^{\mu\nu}\Big)
-k^\lambda\Big(\big(q_1^2-q_2^2\big)\big(q_1^\mu+q_2^\mu\big)\big(q_1^\nu+q_2^\nu\big)-s\big(q_1^\mu q_1^\nu-q_2^\mu q_2^\nu\big)\Big)\bigg]\notag\\
&+\bigg[2\big(q_1^\lambda+q_2^\lambda\big)q_{tu}^\sigma-\big(s-q_3^2\big)\big(q_1^\lambda q_2^\sigma-q_2^\lambda q_1^\sigma\big)+2\lam \bar z_s k^\lambda q_3^\sigma\bigg]\notag\\
&\qquad\times\bigg[\big(s-q_1^2+q_2^2\big)\big(q_1^\mu k^\nu+k^\mu q_1^\nu\big)-\big(s+q_1^2-q_2^2\big)\big(q_2^\mu k^\nu+k^\mu q_2^\nu\big)\bigg]\notag\\
&\hspace{-2pt}+2k^\lambda\bigg[s\Big(k^\nu\big(q_1^\mu+q_2^\mu\big)+k^\mu\big(q_1^\nu+q_2^\nu\big)\Big)\big(q_2^2\,q_1^\sigma+q_1^2\,q_2^\sigma\big)
-s^2\Big(q_1^\sigma\big(q_2^\mu k^\nu+k^\mu q_2^\nu\big)+q_2^\sigma\big(q_1^\mu k^\nu+k^\mu q_1^\nu\big)\Big)\notag\\
&\qquad-q_3^\sigma\Big(\big(\lam+s q_2^2\big)\big(q_1^\mu k^\nu+k^\mu q_1^\nu\big)+\big(\lam+s q_1^2\big)\big(q_2^\mu k^\nu+k^\mu q_2^\nu\big)\Big)\bigg]\notag\\
&+\frac{2\lam}{s-q_3^2}k^\mu k^\nu q_3^\sigma\Big(\big(s-q_3^2\big)q_3^\lambda+\big(s+q_3^2\big)k^\lambda\Big)\Bigg\},\notag\\
A_{7,s}^{\mu\nu\lambda\sigma}&=\frac{16}{15\big(s-q_3^2\big)^2\lam}\Big(k\cdot q_3\,q_3^\lambda-q_3^2\,k^\lambda\Big)\Bigg\{
\frac{\lam}{s}\Big(k^\mu q_3^\sigma-k\cdot q_3\,g^{\mu\sigma}\Big)\bigg(\frac{s\, k^\nu}{s-q_3^2}+k_s^\nu\bigg)\notag\\
&+\frac{\lam}{s}\Big(k^\nu q_3^\sigma-k\cdot q_3\,g^{\nu\sigma}\Big)\bigg(\frac{s\, k^\mu}{s-q_3^2}+k_s^\mu\bigg)\notag\\
&-\frac{2q_{tu}^\sigma}{s-q_3^2}\Big(\big(s-q_1^2+q_2^2\big)\big(q_1^\mu k^\nu+k^\mu q_1^\nu\big)-\big(s+q_1^2-q_2^2\big)\big(q_2^\mu k^\nu+k^\mu q_2^\nu\big)\Big)\notag\\
&+q_{tu}^\sigma\bigg[\lam \bar z_s g^{\mu\nu}+2\big(q_1^\mu q_1^\nu-q_2^\mu q_2^\nu\big)-\frac{2}{s}\big(q_1^\mu+q_2^\mu\big)\big(q_1^\nu+q_2^\nu\big)\big(q_1^2-q_2^2+\lam \bar z_s\big)-4\bar z_s q_{12}^{\mu\nu}\bigg]
\Bigg\},\notag\\
A_{8,s}^{\mu\nu\lambda\sigma}&=\frac{32\bar z_s}{5\big(s-q_3^2\big)^2\lam}\Big(q_2^2\,q_1^\nu-q_1\cdot q_2\, q_2^\nu\Big)
\Big(k^\lambda q_3^\sigma-k\cdot q_3\,g^{\lambda\sigma}\Big)\notag\\
&\times\bigg\{\big(t-q_2^2\big)\big(q_1\cdot q_2\,q_1^\mu-q_1^2\,q_2^\mu\big)
-\big(u-q_1^2\big)\big(q_2^2\,q_1^\mu-q_1\cdot q_2\,q_2^\mu\big)+\frac{\lam}{2}k^\mu\bigg\},\notag\\
A_{9,s}^{\mu\nu\lambda\sigma}&=\frac{16}{15\big(s-q_3^2\big)\lam}\Big(q_2^2\,q_1^\nu-q_1\cdot q_2\, q_2^\nu\Big)\Bigg\{
\frac{2q_{tu}^\sigma}{s-q_3^2}\Big(q_3^\mu k^\lambda-k\cdot q_3\, g^{\mu\lambda}\Big)+g^{\mu\sigma}q_{tu}^\lambda\notag\\
&-\frac{4q_{tu}^\sigma q_{tu}^\lambda}{\big(s-q_3^2\big)\lam}\Big(q_1^\mu\big(s-q_1^2+q_2^2\big)-q_2^\mu\big(s+q_1^2-q_2^2\big)\Big)\notag\\
&-\frac{\lam \bar z_s}{s}\Big(k^\lambda q_3^\sigma-k\cdot q_3\, g^{\lambda\sigma}\Big)\bigg[q_1^\mu+q_2^\mu-\bar z_s \Big(q_1^\mu\big(s-q_1^2+q_2^2\big)-q_2^\mu\big(s+q_1^2-q_2^2\big)\Big)\bigg]\notag\\
&+\frac{k^\mu}{s-q_3^2}\bigg[2\lam \bar z_s\Big(k^\lambda q_3^\sigma-k\cdot q_3\,g^{\lambda\sigma}\Big)+2q_{tu}^\sigma\bigg(\frac{2q_3^2\,k^\lambda}{s-q_3^2}+q_3^\lambda\bigg)+2q_3^\sigma q_{tu}^\lambda\bigg]
\Bigg\},\notag\\
A_{10,s}^{\mu\nu\lambda\sigma}&=\frac{32}{15s\big(s-q_3^2\big)^2\lam}\Big(k\cdot q_3\,q_3^\lambda-q_3^2\,k^\lambda\Big)\Big(q_2^2\,q_1^\nu-q_1\cdot q_2\,q_2^\nu\Big)\notag\\
&\times\Bigg\{s\Big(k^\mu q_1^\sigma-k\cdot q_1\,g^{\mu\sigma}-k^\mu q_2^\sigma+k\cdot q_2\, g^{\mu\sigma}\Big)+\big(q_1^2-q_2^2\big)\Big(k^\mu q_3^\sigma-k\cdot q_3\,g^{\mu\sigma}\Big)\notag\\
&\qquad+q_{tu}^\sigma\bigg[\frac{4s\, k^\mu}{s-q_3^2}-q_1^\mu-q_2^\mu+2\bar z_s\Big(q_1^\mu\big(s-q_1^2+q_2^2\big)-q_2^\mu\big(s+q_1^2-q_2^2\big)\Big)\bigg]\Bigg\},\notag\\
A_{11,s}^{\mu\nu\lambda\sigma}&=\frac{32\bar z_s}{5\big(s-q_3^2\big)^2\lam}\Big(q_1\cdot q_2\, q_1^\mu-q_1^2\,q_2^\mu\Big)
\Big(k^\lambda q_3^\sigma-k\cdot q_3\,g^{\lambda\sigma}\Big)\notag\\
&\times\bigg\{\big(t-q_2^2\big)\big(q_1\cdot q_2\,q_1^\nu-q_1^2\,q_2^\nu\big)
-\big(u-q_1^2\big)\big(q_2^2\,q_1^\nu-q_1\cdot q_2\,q_2^\nu\big)+\frac{\lam}{2}k^\nu\bigg\},\notag\\
A_{12,s}^{\mu\nu\lambda\sigma}&=\frac{16}{15\big(s-q_3^2\big)\lam}\Big(q_1\cdot q_2\, q_1^\mu-q_1^2\,q_2^\mu\Big)\Bigg\{
\frac{2q_{tu}^\sigma}{s-q_3^2}\Big(q_3^\nu k^\lambda-k\cdot q_3\, g^{\nu\lambda}\Big)+g^{\nu\sigma}q_{tu}^\lambda\notag\\
&-\frac{4q_{tu}^\sigma q_{tu}^\lambda}{\big(s-q_3^2\big)\lam}\Big(q_1^\nu\big(s-q_1^2+q_2^2\big)-q_2^\nu\big(s+q_1^2-q_2^2\big)\Big)\notag\\
&-\frac{\lam \bar z_s}{s}\Big(k^\lambda q_3^\sigma-k\cdot q_3\, g^{\lambda\sigma}\Big)\bigg[q_1^\nu+q_2^\nu-\bar z_s \Big(q_1^\nu\big(s-q_1^2+q_2^2\big)-q_2^\nu\big(s+q_1^2-q_2^2\big)\Big)\bigg]\notag\\
&+\frac{k^\nu}{s-q_3^2}\bigg[2\lam \bar z_s\Big(k^\lambda q_3^\sigma-k\cdot q_3\,g^{\lambda\sigma}\Big)+2q_{tu}^\sigma\bigg(\frac{2q_3^2\,k^\lambda}{s-q_3^2}+q_3^\lambda\bigg)+2q_3^\sigma q_{tu}^\lambda\bigg]
\Bigg\},\notag\\
A_{13,s}^{\mu\nu\lambda\sigma}&=\frac{32}{15s\big(s-q_3^2\big)^2\lam}\Big(k\cdot q_3\,q_3^\lambda-q_3^2\,k^\lambda\Big)\Big(q_1\cdot q_2\,q_1^\mu-q_1^2\,q_2^\mu\Big)\notag\\
&\times\Bigg\{s\Big(k^\nu q_1^\sigma-k\cdot q_1\,g^{\nu\sigma}-k^\nu q_2^\sigma+k\cdot q_2\, g^{\nu\sigma}\Big)+\big(q_1^2-q_2^2\big)\Big(k^\nu q_3^\sigma-k\cdot q_3\,g^{\nu\sigma}\Big)\notag\\
&\qquad+q_{tu}^\sigma\bigg[\frac{4s\, k^\nu}{s-q_3^2}-q_1^\nu-q_2^\nu+2\bar z_s\Big(q_1^\nu\big(s-q_1^2+q_2^2\big)-q_2^\nu\big(s+q_1^2-q_2^2\big)\Big)\bigg]\Bigg\},\notag\\
A_{14,s}^{\mu\nu\lambda\sigma}&=-\frac{8}{5\big(s-q_3^2\big)\lam}\bigg\{\big(1-z_s^2\big)\Big(k^\lambda q_3^\sigma-k\cdot q_3\,g^{\lambda\sigma}\Big)
+\frac{4s\,q_{tu}^\lambda q_{tu}^\sigma}{\big(s-q_3^2\big)\lam}\bigg\}\notag\\
&\times\Big(q_1\cdot q_2\,q_1^\mu-q_1^2\,q_2^\mu\Big)\Big(q_2^2\,q_1^\nu-q_1\cdot q_2\,q_2^\nu\Big),\notag\\
A_{15,s}^{\mu\nu\lambda\sigma}&=\frac{32\bar z_s\,q_{tu}^\sigma}{5\big(s-q_3^2\big)^2\lam}\Big(k\cdot q_3\,q_3^\lambda-q_3^2\,k^\lambda\Big)
\Big(q_1\cdot q_2\,q_1^\mu-q_1^2\,q_2^\mu\Big)\Big(q_2^2\,q_1^\nu-q_1\cdot q_2\,q_2^\nu\Big),
\end{align}
where $\bar z_s=z_s/\sqrt{\lam}$, $z_s$ is defined in~\eqref{schannel_angle}, and
\begin{align}
 q_{tu}^\lambda&=\big(t-q_2^2\big)q_1^\lambda-\big(u-q_1^2\big)q_2^\lambda-\lam \bar z_s\,k^\lambda,\qquad
 q_{tu}^\sigma=\big(t-q_2^2\big)q_1^\sigma-\big(u-q_1^2\big)q_2^\sigma,\notag\\
 k_s^\mu&=\frac{s\, k^\mu}{s-q_3^2}-q_1^\mu\Big(1-\bar z_s\big(s-q_1^2+q_2^2\big)\Big)
 -q_2^\mu\Big(1+\bar z_s\big(s+q_1^2-q_2^2\big)\Big),\notag\\
 q^{\mu\nu}_{12}&=q_1^\mu\Big(q_2^2\,q_1^\nu-q_1\cdot q_2\,q_2^\nu\Big)
-q_2^\mu\Big(q_1\cdot q_2\,q_1^\nu-q_1^2\,q_2^\nu\Big).
\end{align}

\section{Unitarity and helicity amplitudes}
\label{app:unitarity_hel}

The imaginary parts of the helicity amplitudes that follow from $\Pi^{\mu\nu\lambda\sigma}$ by contraction with the pertinent polarization vectors have to reproduce the imaginary parts as expected from general arguments about helicity amplitudes~\cite{JW}.
In our conventions we find for momenta and polarization vectors
\begin{align}
\label{lbl_pol_vec}
 q_1^\mu&=\big(E_1,0,0,q_{12}\big), &
 q_3^\mu&=\big(-E_3,-q_{34}\sin\theta_s,0,-q_{34}\cos\theta_s\big),\notag\\
  q_2^\mu&=\big(E_2,0,0,-q_{12}\big), &
 k^\mu&=\big(E_4,-q_{34}\sin\theta_s,0,-q_{34}\cos\theta_s\big),\notag\\
 \epsilon^{\mu}\big(q_1,\pm\big)&=\mp\frac{1}{\sqrt{2}}\big(0,1,\pm i,0\big),&
  \epsilon^{\mu*}\big(-q_3,\pm\big)&=\mp\frac{1}{\sqrt{2}}\big(0,\cos\theta_s,\mp i,-\sin\theta_s\big),\notag\\
 \epsilon^{\mu}\big(q_2,\pm\big)&=\mp\frac{1}{\sqrt{2}}\big(0,1,\mp i,0\big),&
 \epsilon^{\mu*}\big(k,\pm\big)&=\mp\frac{1}{\sqrt{2}}\big(0,\cos\theta_s,\pm i,-\sin\theta_s\big),\notag\\
 \epsilon^\mu\big(q_1,0\big)&=\frac{1}{\xi_1}\big(q_{12},0,0,E_1\big),&
 \epsilon^{\mu*}\big(-q_3,0\big)&=\frac{1}{\xi_3}\big(q_{34},E_3\sin\theta_s,0,E_3\cos\theta_s\big),\notag\\
  \epsilon^\mu\big(q_2,0\big)&=\frac{1}{\xi_2}\big(-q_{12},0,0,E_2\big),
\end{align}
where
\beq
E_1=\frac{s+q_1^2-q_2^2}{2\sqrt{s}},\qquad E_2=\frac{s-q_1^2+q_2^2}{2\sqrt{s}},\qquad E_3=\frac{s+q_3^2}{2\sqrt{s}},\qquad
E_4=\frac{s-q_3^2}{2\sqrt{s}},
\eeq
and
\beq
q_{12}=\frac{\sqrt{\lambda_{12}}}{2\sqrt{s}},\qquad q_{34}=\frac{s-q_3^2}{2\sqrt{s}}.
\eeq
Contracting these expressions with the $A_{i,s}^{\mu\nu\lambda\sigma}$ from App.~\ref{app:Lorentz}, we find the following imaginary parts\footnote{The additional sign for each occurrence of the amplitude $h_{0+}$ originates from our convention in~\eqref{def_partial_waves}, since $d^J_{10}=-d^J_{-1,0}$.}
\begin{align}
\label{hel_im_part}
 \Imspipi H_{++,++}&=\frac{\sigma_s}{16\pi}\Big\{h_{0,++}\big(s;q_1^2,q_2^2\big)h_{0,++}\big(s;q_3^2,0\big)
 +5P_2\big(z_s\big)h_{2,++}\big(s;q_1^2,q_2^2\big)h_{2,++}\big(s;q_3^2,0\big)\Big\},\notag\\
 \Imspipi H_{00,++}&=\frac{\sigma_s}{16\pi}\Big\{h_{0,00}\big(s;q_1^2,q_2^2\big)h_{0,++}\big(s;q_3^2,0\big)
 +5P_2\big(z_s\big)h_{2,00}\big(s;q_1^2,q_2^2\big)h_{2,++}\big(s;q_3^2,0\big)\Big\},\notag\\
 \Imspipi H_{++,+-}&=\frac{\sigma_s}{16\pi}\,5\,d^2_{02}\big(z_s\big)h_{2,++}\big(s;q_1^2,q_2^2\big)h_{2,+-}\big(s;q_3^2,0\big),\notag\\
  \Imspipi H_{++,0+}&=-\frac{\sigma_s}{16\pi}\,5\,d^2_{0,-1}\big(z_s\big)h_{2,++}\big(s;q_1^2,q_2^2\big)h_{2,0+}\big(s;q_3^2,0\big),\notag\\
  \Imspipi H_{+-,++}&=\frac{\sigma_s}{16\pi}\,5\,d^2_{20}\big(z_s\big)h_{2,+-}\big(s;q_1^2,q_2^2\big)h_{2,++}\big(s;q_3^2,0\big),\notag\\
  \Imspipi H_{+-,+-}&=\frac{\sigma_s}{16\pi}\,5\,d^2_{22}\big(z_s\big)h_{2,+-}\big(s;q_1^2,q_2^2\big)h_{2,+-}\big(s;q_3^2,0\big),\notag\\
  \Imspipi H_{+-,0+}&=-\frac{\sigma_s}{16\pi}\,5\,d^2_{2,-1}\big(z_s\big)h_{2,+-}\big(s;q_1^2,q_2^2\big)h_{2,0+}\big(s;q_3^2,0\big),\notag\\
  \Imspipi H_{+0,++}&=\frac{\sigma_s}{16\pi}\,5\,d^2_{10}\big(z_s\big)h_{2,+0}\big(s;q_1^2,q_2^2\big)h_{2,++}\big(s;q_3^2,0\big),\notag\\
  \Imspipi H_{+0,+-}&=\frac{\sigma_s}{16\pi}\,5\,d^2_{12}\big(z_s\big)h_{2,+0}\big(s;q_1^2,q_2^2\big)h_{2,+-}\big(s;q_3^2,0\big),\notag\\
  \Imspipi H_{+0,0+}&=-\frac{\sigma_s}{16\pi}\,5\,d^2_{1,-1}\big(z_s\big)h_{2,+0}\big(s;q_1^2,q_2^2\big)h_{2,0+}\big(s;q_3^2,0\big),\notag\\
  \Imspipi H_{0+,++}&=-\frac{\sigma_s}{16\pi}\,5\,d^2_{-1,0}\big(z_s\big)h_{2,0+}\big(s;q_1^2,q_2^2\big)h_{2,++}\big(s;q_3^2,0\big),\notag\\
  \Imspipi H_{0+,+-}&=-\frac{\sigma_s}{16\pi}\,5\,d^2_{-1,2}\big(z_s\big)h_{2,0+}\big(s;q_1^2,q_2^2\big)h_{2,+-}\big(s;q_3^2,0\big),\notag\\
  \Imspipi H_{0+,0+}&=\frac{\sigma_s}{16\pi}\,5\,d^2_{-1,-1}\big(z_s\big)h_{2,0+}\big(s;q_1^2,q_2^2\big)h_{2,0+}\big(s;q_3^2,0\big),\notag\\
  \Imspipi H_{00,+-}&=\frac{\sigma_s}{16\pi}\,5\,d^2_{02}\big(z_s\big)h_{2,00}\big(s;q_1^2,q_2^2\big)h_{2,+-}\big(s;q_3^2,0\big),\notag\\
  \Imspipi H_{00,0+}&=-\frac{\sigma_s}{16\pi}\,5\,d^2_{0,-1}\big(z_s\big)h_{2,00}\big(s;q_1^2,q_2^2\big)h_{2,0+}\big(s;q_3^2,0\big).
\end{align}
Indeed, these expressions could have been written down immediately based on
general properties of helicity partial waves~\cite{JW}, and thus provide a
powerful check on the calculation of the
$A_{i,s}^{\mu\nu\lambda\sigma}$. Moreover, they provide the proof that the
general decomposition of the HLbL tensor in~\eqref{lbl_tensor} leads to
diagonal unitarity relations. 

\section{Scalar Roy equations}
\label{app:roy}

Dispersion relations for single-variable functions can be constructed in
close analogy to $\pi\pi$ Roy equations~\cite{Roy}. We illustrate this here
for a scalar example, e.g.\ $\pi\pi$ scattering without isospin. The
starting point in the derivation is given by a twice-subtracted fixed-$t$
dispersion relation for the scattering amplitude $T(s,t)$ 
\beq
T(s,t)=C(t)+\frac{1}{\pi}\int\limits_{4\mpi^2}^\infty \frac{\diff s'}{s'^2}\bigg\{\frac{s^2}{s'-s}+\frac{u^2}{s'-u}\bigg\}\Im T(s',t).
\eeq
The subtraction function $C(t)$ can be determined by imposing $st$ crossing
symmetry in the form $T(0,t)=T(t,0)$, leading to 
\begin{align}
\label{Roy_der}
 T(s,t)&=C(0)+\frac{1}{\pi}\int\limits_{4\mpi^2}^\infty \frac{\diff s'}{s'^2}\bigg\{\frac{s^2}{s'-s}
 +\frac{t^2}{s'-t}+\frac{u^2}{s'-u}\bigg\}\Im T(s',t)\notag\\
 &+\frac{1}{\pi}\int\limits_{4\mpi^2}^\infty \frac{\diff s'}{s'^2}\bigg\{\frac{t^2}{s'-t}
 +\frac{\big(4\mpi^2-t\big)^2}{s'-4\mpi^2+t}\bigg\}\Big[\Im T(s',0)-\Im T(s',t)\Big].
\end{align}
Due to Bose symmetry only even partial waves are allowed in the absence of isospin, so that restricting ourselves to the $S$-wave $t_0(s)$, \eqref{Roy_der} becomes
\begin{align}
\label{Swave_scalar}
T(s,t)&=C(0)+\frac{1}{\pi}\int\limits_{4\mpi^2}^\infty \frac{\diff s'}{s'^2}\bigg\{\frac{s^2}{s'-s}
 +\frac{t^2}{s'-t}+\frac{u^2}{s'-u}\bigg\}\Im t_0(s')+(l\geq 2)\notag\\
 &=T(s)+T(t)+T(u)+(l\geq 2),\qquad 
 T(s)=\frac{C(0)}{3}+\frac{s^2}{\pi}\int\limits_{4\mpi^2}^\infty \frac{\diff s'\,\Im t_0(s')}{s'^2(s'-s)},
\end{align}
and the amplitude factorizes into single-variable functions
$T(s)$.\footnote{A similar decomposition has been used for a dispersive
  description of the processes
  $\gamma\pi\to\pi\pi$~\cite{Hannah,Truong,HKS} and $\omega,\phi\to
  3\pi$~\cite{NKS}, where only odd partial waves are allowed.  
In the $P$-wave approximation one finds a result completely analogous
to~\eqref{Swave_scalar}. The extension to the $F$-wave is discussed
in~\cite{NKS}.} For HLbL scattering we encounter precisely the same
situation that only the leading partial wave in a given amplitude is
relevant. Moreover, the analog of the parameter $C(0)$ is determined by
soft-photon constraints, whose precise form can be inferred from the kernel
functions in~\eqref{kernels_diagonal}. 

\section{Angular average}
\label{app:average}

The $D$-wave contributions involve terms such as $k\cdot q_i/k\cdot q_3$, whose 
limit for $k\to 0$ depends on the direction in which $k$ is
taken to zero.
Therefore, even though all $A_{i}^{\mu\nu\lambda\sigma}$ scale as $\Order(1)$,
the result for the derivative in the limit $k\to0$ is ambiguous.
Such terms also appear in $\gamma^*\gamma^*\to\pi\pi$, see~\eqref{helicity_amplitudes}, e.g.\ $H_{++}$ involves a $D$-wave contribution 
without the expected angular-momentum factor for the photon pair, so that the same phenomenon occurs for $q_t\to 0$~\cite{Moussallam13,CHPS_prep}.
A generalization of~\eqref{lblmaster}
valid also in this case may be derived by including these terms in the average over the spatial directions
of $k$
\begin{align}
\label{lbl_master_average}
 a_\mu&=\frac{1}{16m}\Tr\Big\{\big(\slashed{p}+m\big)\big[\gamma^\rho,\gamma^\tau\big]\big(\slashed{p}+m\big)\tilde\Gamma_{\rho\tau}\Big\},\\  
 \tilde\Gamma_{\rho\tau}&=-e^6\int\frac{\diff^4q_1}{(2\pi)^4}\int\frac{\diff^4q_2}{(2\pi)^4}\frac{1}{q_1^2q_2^2s}
 \frac{\gamma^\mu\big(\slashed{p}+\slashed{q_1}+m\big)\gamma^\lambda\big(\slashed{p}-\slashed{q_2}+m)\gamma^\nu}{\big((p+q_1)^2-m^2\big)
 \big((p-q_2)^2-m^2\big)}\bigg[\int\frac{\diff \Omega(p,k)}{4\pi}\frac{k_\tau k^\sigma}{k^2}\frac{\partial}{\partial k^\rho}\Pi_{\mu\nu\lambda\sigma}\bigg]_{k=0},\notag
\end{align}
where the angular average occurs with
respect to the fixed axis defined by $p$. The tensor
decomposition 
\beq
\bigg[\int\frac{\diff \Omega(p,k)}{4\pi}\frac{k^\mu k^\nu}{k^2}\bigg]_{k=0}=\frac{1}{3}\Sigma^{\mu\nu},\qquad \Sigma^{\mu\nu}=g^{\mu\nu}-\frac{p^\mu p^\nu}{m^2},
\eeq
then reproduces~\eqref{lblmaster}, as the terms depending on $p$ vanish in
the trace.

Here, we need a generalization involving the following integrals
\begin{align}
 \bigg[\int\frac{\diff \Omega(p,k)}{4\pi}\frac{k^\mu k^\nu k^\lambda}{k^2\,k\cdot q_3}\bigg]_{k=0}&=\frac{m^2}{3Z}\bigg[
 \xi_p\Big(\Sigma^{\mu\nu}p^\lambda+\Sigma^{\mu\lambda}p^\nu+\Sigma^{\nu\lambda}p^\mu\Big)
 -\Big(\Sigma^{\mu\nu}q_3^\lambda+\Sigma^{\mu\lambda}q_3^\nu+\Sigma^{\nu\lambda}q_3^\mu\Big)\bigg]\notag\\
 &\hspace{-28pt}+\frac{2m^4}{3Z^2}\bigg[\xi_p^3p^\mu p^\nu p^\lambda-\xi_p^2\Big(p^\mu p^\nu q_3^\lambda+p^\mu p^\lambda q_3^\nu+p^\nu p^\lambda q_3^\mu\Big)\notag\\
 &\hspace{-28pt}\quad+\xi_p\Big(q_3^\mu q_3^\nu p^\lambda+q_3^\mu q_3^\lambda p^\nu+q_3^\nu q_3^\lambda p^\mu\Big)-q_3^\mu q_3^\nu q_3^\lambda\bigg],\notag\\
 \bigg[\int\frac{\diff \Omega(p,k)}{4\pi}\frac{k^\mu k^\nu k^\lambda k^\sigma}{k^2\,(k\cdot q_3)^2}\bigg]_{k=0}&=\frac{m^2}{3Z}\Big(\Sigma^{\mu\nu}\Sigma^{\lambda\sigma}+\Sigma^{\mu\lambda}\Sigma^{\nu\sigma}+\Sigma^{\mu\sigma}\Sigma^{\nu\lambda}\Big)\notag\\
 &\hspace{-28pt}+\frac{2m^4}{3Z^2}\bigg[\xi_p^2\Big(\Sigma^{\mu\nu}p^\lambda p^\sigma+\Sigma^{\mu\lambda}p^\nu p^\sigma+\Sigma^{\mu\sigma}p^\nu p^\lambda+\Sigma^{\nu\lambda}p^\mu p^\sigma+\Sigma^{\nu\sigma}p^\mu p^\lambda+\Sigma^{\lambda\sigma}p^\mu p^\nu\Big)\notag\\
 &\hspace{-28pt}\quad-\xi_p\Big(\Sigma^{\mu\nu}\big(p^\lambda q_3^\sigma+p^\sigma q_3^\lambda\big)
 +\Sigma^{\mu\lambda}\big(p^\nu q_3^\sigma+p^\sigma q_3^\nu\big)+\Sigma^{\mu\sigma}\big(p^\nu q_3^\lambda+p^\lambda q_3^\nu\big)\notag\\
 &\hspace{-28pt}\quad\quad+\Sigma^{\nu\lambda}\big(p^\mu q_3^\sigma+p^\sigma q_3^\mu\big)+\Sigma^{\nu\sigma}\big(p^\mu q_3^\lambda+p^\lambda q_3^\mu\big)+\Sigma^{\lambda\sigma}\big(p^\mu q_3^\nu+p^\nu q_3^\mu\big)\Big)\notag\\
 &\hspace{-28pt}\quad+\Big(\Sigma^{\mu\nu}q_3^\lambda q_3^\sigma+\Sigma^{\mu\lambda}q_3^\nu q_3^\sigma+\Sigma^{\mu\sigma}q_3^\nu q_3^\lambda+\Sigma^{\nu\lambda}q_3^\mu q_3^\sigma+\Sigma^{\nu\sigma}q_3^\mu q_3^\lambda+\Sigma^{\lambda\sigma}q_3^\mu q_3^\nu\Big)\bigg]\notag\\
 &\hspace{-28pt}+\frac{8m^6}{3Z^3}\bigg[\xi_p^4p^\mu p^\nu p^\lambda p^\sigma
 -\xi_p^3\Big(p^\mu p^\nu p^\lambda q_3^\sigma+p^\mu p^\nu p^\sigma q_3^\lambda+p^\mu p^\lambda p^\sigma q_3^\nu+p^\nu p^\lambda p^\sigma q_3^\mu\Big)\notag\\
 &\hspace{-28pt}\quad+\xi_p^2\Big(p^\mu p^\nu q_3^\lambda q_3^\sigma+p^\mu p^\lambda q_3^\nu q_3^\sigma+p^\mu p^\sigma q_3^\nu q_3^\lambda+p^\nu p^\lambda q_3^\mu q_3^\sigma+p^\nu p^\sigma q_3^\mu q_3^\lambda+p^\lambda p^\sigma q_3^\mu q_3^\nu\Big)\notag\\
 &\hspace{-28pt}\quad-\xi_p\Big(q_3^\mu q_3^\nu q_3^\lambda p^\sigma+q_3^\mu q_3^\nu q_3^\sigma p^\lambda+q_3^\mu q_3^\lambda q_3^\sigma p^\nu+q_3^\nu q_3^\lambda q_3^\sigma p^\mu\Big)+q_3^\mu q_3^\nu q_3^\lambda q_3^\sigma\bigg],
\end{align}
where
\beq
Z=(p\cdot q_3)^2-m^2q_3^2,\qquad \xi_p=\frac{p\cdot q_3}{m^2}.
\eeq
The result for the fourth-order tensor can most easily be obtained by means
of 
\beq
 \int\frac{\diff \Omega(p,k)}{4\pi}\frac{k^\mu k^\nu k^\lambda
   k^\sigma}{k^2\,(k\cdot q_3)^2}=-\frac{\partial}{\partial
   q_{3\sigma}}\int\frac{\diff \Omega(p,k)}{4\pi}\frac{k^\mu k^\nu
   k^\lambda}{k^2\,k\cdot q_3}. 
\eeq
A powerful check on the calculation is provided by gauge invariance, as the
result after the angular average still has to vanish when contracted with
$q_1^\mu$, $q_2^\nu$, or $(q_1+q_2)^\lambda$.

\section{Dispersion integrals}
\label{app:master_int}

The dispersive integrals in~\eqref{final_master} read (including only diagonal kernels for $D$-waves)
\begin{align}
 I_{1,s}&=\frac{1}{\pi}\int\limits_{4\mpi^2}^\infty\frac{\diff s'}{s'-s}\bigg[\bigg(\frac{1}{s'-s}
 -\frac{s'-q_1^2-q_2^2}{\lambda\big(s',q_1^2,q_2^2\big)}\bigg)\Im \bar h_{++,++}^0\big(s';q_1^2,q_2^2;s,0\big)\\
 &\qquad+\frac{2\xi_1\xi_2}{\lambda\big(s',q_1^2,q_2^2\big)} \Im \bar h^0_{00,++}\big(s';q_1^2,q_2^2;s,0\big)\bigg],\notag\\
I_{1,u}&=\frac{1}{\pi}\int\limits_{4\mpi^2}^\infty\frac{\diff s'}{s'-q_1^2}\bigg[\bigg(\frac{1}{s'-q_1^2}
 -\frac{s'-s-q_2^2}{\lambda\big(s',s,q_2^2\big)}\bigg)\Im \bar h_{++,++}^0\big(s';s,q_2^2;q_1^2,0\big)\notag\\
 &\qquad+\frac{2\xi_s\xi_2}{\lambda\big(s',s,q_2^2\big)} \Im \bar h^0_{00,++}\big(s';s,q_2^2;q_1^2,0\big)\bigg],\notag\\
  I_{2,s}&=\frac{1}{\pi}\int\limits_{4\mpi^2}^\infty\frac{\diff s'}{s'-s}\bigg[\bigg(\frac{1}{s'-s}
 -\frac{s'-q_1^2-q_2^2}{\lambda\big(s',q_1^2,q_2^2\big)}\bigg)\Im \bar h_{00,++}^0\big(s';q_1^2,q_2^2;s,0\big)\notag\\
 &\qquad+\frac{2q_1^2q_2^2}{\xi_1\xi_2\lambda\big(s',q_1^2,q_2^2\big)}\Im \bar h^0_{++,++}\big(s';q_1^2,q_2^2;s,0\big)\bigg]\bigg(-\frac{\xi_1\xi_2}{q_1^2q_2^2}\bigg),\notag\\
I_{2,u}&=\frac{1}{\pi}\int\limits_{4\mpi^2}^\infty\frac{\diff s'}{s'-q_1^2}\bigg[\bigg(\frac{1}{s'-q_1^2}
 -\frac{s'-s-q_2^2}{\lambda\big(s',s,q_2^2\big)}\bigg)\Im \bar h_{00,++}^0\big(s';s,q_2^2;q_1^2,0\big)\notag\\
&\qquad+\frac{2s\,q_2^2}{\xi_s\xi_2\lambda\big(s',s,q_2^2\big)}\Im \bar h^0_{++,++}\big(s';s,q_2^2;q_1^2,0\big)\bigg] 
 \bigg(-\frac{\xi_s\xi_2}{s\,q_2^2}\bigg),\notag\\
   I_{3,s}&=\frac{1}{\pi}\int\limits_{4\mpi^2}^\infty\frac{\diff s'\, s'}{\big(s'-s\big)\lambda\big(s',q_1^2,q_2^2\big)}\bigg(\frac{1}{s'-s}
 -\frac{s'-q_1^2-q_2^2}{\lambda\big(s',q_1^2,q_2^2\big)}\bigg)\Im \bar h_{++,+-}^2\big(s';q_1^2,q_2^2;s,0\big)\bigg(-\frac{25}{4}\sqrt{6}\bigg),\notag\\
  I_{3,u}&=\frac{1}{\pi}\int\limits_{4\mpi^2}^\infty\frac{\diff s'\, s'}{\big(s'-q_1^2\big)\lambda\big(s',s,q_2^2\big)}\bigg(\frac{1}{s'-q_1^2}
 -\frac{s'-s-q_2^2}{\lambda\big(s',s,q_2^2\big)}\bigg)\Im \bar h_{++,+-}^2\big(s';s,q_2^2;q_1^2,0\big)\bigg(-\frac{25}{4}\sqrt{6}\bigg),\notag\\
  I_{6,s}&=\frac{1}{\pi}\int\limits_{4\mpi^2}^\infty\frac{\diff s'}{\big(s'-q_1^2-q_2^2\big)\big(s'-s\big)^2}
  \Im \bar h_{+-,+-}^2\big(s';q_1^2,q_2^2;s,0\big)\bigg(\frac{75}{8}\bigg),\notag\\
  I_{6,u}&=\frac{1}{\pi}\int\limits_{4\mpi^2}^\infty\frac{\diff s'}{\big(s'-s-q_2^2\big)\big(s'-q_1^2\big)^2}
  \Im \bar h_{+-,+-}^2\big(s';s,q_2^2;q_1^2,0\big)\bigg(\frac{75}{8}\bigg),\notag\\
  I_{9,s}&=\frac{1}{\pi}\int\limits_{4\mpi^2}^\infty\frac{\diff s'\,\sqrt{s'}}{\lambda\big(s',q_1^2,q_2^2\big)\big(s'-s\big)^2}
  \Im \bar h_{+0,+-}^2\big(s';q_1^2,q_2^2;s,0\big)\bigg(-\frac{75}{8}\sqrt{2}\,\frac{\xi_2}{q_2^2}\bigg),\notag\\
  I_{9,u}&=\frac{1}{\pi}\int\limits_{4\mpi^2}^\infty\frac{\diff s'\,\sqrt{s'}}{\lambda\big(s',s,q_2^2\big)\big(s'-q_1^2\big)^2}
  \Im \bar h_{+0,+-}^2\big(s';s,q_2^2;q_1^2,0\big)\bigg(-\frac{75}{8}\sqrt{2}\,\frac{\xi_2}{q_2^2}\bigg),\notag\\
  I_{12,u}&=\frac{1}{\pi}\int\limits_{4\mpi^2}^\infty\frac{\diff s'\,\sqrt{s'}}{\lambda\big(s',s,q_2^2\big)\big(s'-q_1^2\big)^2}
  \Im \bar h_{0+,+-}^2\big(s';s,q_2^2;q_1^2,0\big)\bigg(-\frac{75}{8}\sqrt{2}\,\frac{\xi_s}{s}\bigg),\notag\\
   I_{14,s}&=\frac{1}{\pi}\int\limits_{4\mpi^2}^\infty\frac{\diff s'\, s'}{\big(s'-s\big)\lambda\big(s',q_1^2,q_2^2\big)}\bigg(\frac{1}{s'-s}
 -\frac{s'-q_1^2-q_2^2}{\lambda\big(s',q_1^2,q_2^2\big)}\bigg)\Im \bar h_{00,+-}^2\big(s';q_1^2,q_2^2;s,0\big)\bigg(\frac{25}{4}\sqrt{6}\,\frac{\xi_1\xi_2}{q_1^2q_2^2}\bigg),\notag\\
  I_{14,u}&=\frac{1}{\pi}\int\limits_{4\mpi^2}^\infty\frac{\diff s'\, s'}{\big(s'-q_1^2\big)\lambda\big(s',s,q_2^2\big)}\bigg(\frac{1}{s'-q_1^2}
 -\frac{s'-s-q_2^2}{\lambda\big(s',s,q_2^2\big)}\bigg)\Im \bar h_{00,+-}^2\big(s';s,q_2^2;q_1^2,0\big)\bigg(\frac{25}{4}\sqrt{6}\,\frac{\xi_s\xi_2}{s\,q_2^2}\bigg),\notag
\end{align}
with the notation
\beq
\Im \bar h_{\lambda_1\lambda_2,\lambda_3\lambda_4}^J\big(s;q_1^2,q_2^2;q_3^2,q_4^2\big)=
\frac{\sigma_s}{16\pi}\theta\big(s-4\mpi^2\big)\,\Sym\Big[h_{J,\lambda_1\lambda_2}\big(s;q_1^2,q_2^2\big)h_{J,\lambda_3\lambda_4}^*\big(s;q_3^2,q_4^2\big)\Big]
\eeq
for the imaginary parts. $\xi_i$ refers to the normalization of the
longitudinal polarization vectors~\eqref{lbl_pol_vec}, $\xi_s$ denotes
$\xi_i$ with $q_i^2\to s$, and $\Sym[\ldots]$ is defined in~\eqref{symmetrizer}.

\section{Integral kernels}
\label{app:master_kernels}

The integration kernels of the final loop integration in~\eqref{final_master} may be expressed as
\begin{align}
 T_{1,s}&=\frac{16}{3} s \bigg\{m^2+\frac{8P_{21}\,p\cdot q_1}{\lambda_{12}}\bigg\},\qquad
 T_{1,u}=\frac{16}{3}\bigg\{\frac{4P_{12}^2}{\lambda_{12}}-P_{12}-Z_u\bigg\},\notag\\
 T_{2,s}&=\frac{128}{3\lambda_{12}}q_2^2\big(P_{12}+P_{21}\big) P_{12},\qquad 
 T_{2,u}=-\frac{8}{3\lambda_{12}}P_{12}\Big\{4P_{21}\big(s+q_1^2-q_2^2\big)+q_2^2\lambda_{12}\Big\},\notag\\
 T_{3,s}&=-\frac{16\big(8P_{21}\,p\cdot q_1+m^2\lambda_{12}\big)}{15\lambda_{12}}\Bigg\{s^2+\big(q_1^2-q_2^2\big)^2+\frac{m^2 s}{Z_s}\Big(\lambda_{12}+4s\big(q_1^2+q_2^2\big)\Big)\notag\\
 &+\frac{s}{Z_s}\bigg(1+\frac{2m^2 s}{Z_s}\bigg)\bigg[m^2\Big(s^2+\big(q_1^2-q_2^2\big)^2\Big)-4s\,p\cdot q_1\,p\cdot q_2-2\big(q_1^2-q_2^2\big)\Big((p\cdot q_1)^2-(p\cdot q_2)^2\Big)\bigg]\Bigg\},\notag\\
 T_{3,u}&=-\frac{32\lambda_{12}m^2P_{12}}{15Z_u}-\frac{16}{15q_1^2}\big(4P_{12}^2-\lambda_{12}Z_u\big)-\frac{16\big(4P_{12}^2-\lambda_{12}(P_{12}+Z_u)\big)}{15\lambda_{12}}
 \Bigg\{4q_2^2+\frac{m^2}{Z_u}\big(8q_1^2q_2^2+\lambda_{12}\big)\notag\\
 &+\frac{4}{Z_u}\bigg(1+\frac{2m^2q_1^2}{Z_u}\bigg)\Big[m^2(q_1\cdot q_2)^2-P_{12}\, p\cdot q_2-p\cdot q_1\,p\cdot q_2\, q_1\cdot q_2\Big]\Bigg\},\notag\\
 T_{6,s}&=\frac{128(P_{12}+P_{21})(s-q_1^2-q_2^2)}{45\lambda_{12}^2}\Bigg\{\lambda_{12}^2+\frac{3\lambda_{12}m^2s}{Z_s}p\cdot q_1\big(s-q_1^2+q_2^2\big)\notag\\
 &+\bigg(1+\frac{3m^2 s}{Z_s}+\frac{2m^4s^2}{Z_s^2}\bigg)p\cdot q_1\big(q_1^2-q_2^2\big)^2\big(3s-q_1^2+q_2^2\big)\notag\\
 &+\frac{s^2}{Z_s}\bigg(1+\frac{2m^2s}{Z_s}\bigg)p\cdot q_1\big(p\cdot q_1-p\cdot q_2\big)^2\big(s-3q_1^2+3q_2^2\big)\Bigg\}-\frac{32}{45}\big(s-q_1^2-q_2^2\big)\big(5m^2s+Z_s\big)
 ,\notag\\
 T_{6,u}&=-\frac{32P_{12}(s-q_1^2+q_2^2)}{45\lambda_{12}^2}\Bigg\{\frac{3\lambda_{12}m^2q_1^2}{Z_u}\big(4P_{12}-\lambda_{12}\big)\notag\\
 &+16\bigg(1+\frac{3m^2q_1^2}{Z_u}+\frac{2m^4q_1^4}{Z_u^2}\bigg)(q_1\cdot q_2)^2\big(P_{12}-2q_1^2\,p\cdot q_2\big)\notag\\
 &+\frac{16q_1^4}{Z_u}\bigg(1+\frac{2m^2q_1^2}{Z_u}\bigg)(p\cdot q_2)^2\big(2q_1^2\,p\cdot q_2+3P_{12}\big)\Bigg\}+\frac{32}{45}\big(s-q_1^2+q_2^2\big)\big(5m^2q_1^2+Z_u\big),\notag\\
 T_{9,s}&=\frac{16(P_{12}+P_{21})}{45\lambda_{12}}\Bigg\{3\lambda_{12}q_2^2\big(s-3q_1^2-q_2^2\big)+8\big(q_1^2-q_2^2\big)^2p\cdot q_2\big(s-q_1^2-3q_2^2\big)\notag\\
 &-2\lambda_{12}\,p\cdot q_2\bigg(s+q_1^2-5q_2^2+\frac{3m^2}{Z_s}\big(q_1^2-q_2^2\big)\big(s-q_1^2+5q_2^2\big)\bigg)\notag\\
 &+\frac{12m^2}{Z_s}\bigg(1+\frac{2m^2 s}{3Z_s}\bigg)p\cdot q_2\big(q_1^2-q_2^2\big)^2\Big(s\big(s-4q_2^2\big)-\big(q_1^2-q_2^2\big)^2\Big)\notag\\
 &+\frac{4s}{Z_s}\bigg(1+\frac{2m^2 s}{Z_s}\bigg)p\cdot q_2\big(p\cdot q_1-p\cdot q_2\big)^2\Big(s\big(s+4q_2^2\big)-\big(q_1^2-q_2^2\big)^2\Big)\notag\\
 &-\frac{3\lambda_{12}q_2^2}{Z_s}\big(s+q_1^2-q_2^2\big)\Big(m^2\big(q_1^2-q_2^2\big)-(p\cdot q_1)^2+(p\cdot q_2)^2\Big)\notag\\
 &+\big(s-q_1^2+q_2^2\big)\big(p\cdot q_1-p\cdot q_2)\bigg[3\lambda_{12}
 +\frac{3\lambda_{12}m^2}{Z_s}\big(s+4q_2^2\big)-4\big(q_1^2-q_2^2\big)\big(s+q_1^2-5q_2^2\big)\bigg]\notag\\
 &-\frac{12m^2s}{Z_s}\bigg(1+\frac{2m^2 s}{3Z_s}\bigg)\big(q_1^2-q_2^2\big)\big(s+q_1^2-5q_2^2\big)\big(s-q_1^2+q_2^2\big)\big(p\cdot q_1-p\cdot q_2\big)\notag\\
 &-\frac{1}{Z_s}\big(s-q_1^2+q_2^2\big)\big(p\cdot q_1-p\cdot q_2\big)^2\big(p\cdot q_1+p\cdot q_2\big)\notag\\
 &\qquad\qquad\times\bigg[\bigg(1+\frac{2m^2 s}{Z_s}\bigg)\Big(s\big(3s+14q_2^2-10q_1^2\big)-\big(q_1^2-q_2^2\big)^2\Big)-\frac{6\lambda_{12}m^2 s}{Z_s}\bigg]\Bigg\}\notag\\
 &-\frac{16}{45}\lambda_{12}\big(2m^2q_2^2+q_2^2\,p\cdot q_1+2(p\cdot q_2)^2\big),\notag\\
 T_{9,u}&=\frac{64P_{12}^2}{45\lambda_{12}}\Bigg\{16q_2^2\,q_1\cdot q_2+\frac{3\lambda_{12}q_2^2}{Z_u}\big(2m^2-p\cdot q_1\big)
 +\frac{16m^2q_1^2q_2^2\,q_1\cdot q_2}{Z_u}\bigg(3+\frac{2m^2q_1^2}{Z_u}\bigg)\notag\\
 &-\frac{16}{Z_u}\bigg(1+\frac{2m^2q_1^2}{Z_u}\bigg)p\cdot q_2\Big(q_1^2q_2^2\,p\cdot q_1+P_{12}\,q_1\cdot q_2\Big)\Bigg\}-\frac{16}{45}q_2^2\lambda_{12}\big(2m^2-p\cdot q_1\big),\notag\\
 T_{12,u}&=\frac{32P_{12}^2}{45\lambda_{12}q_1^2}\Bigg\{\big(s+q_1^2-q_2^2\big)\bigg[8\bigg(1-\frac{4m^4q_1^4}{Z_u^2}\bigg)(q_1\cdot q_2)^2-24q_1^2q_2^2\bigg(1+\frac{2m^2q_1^2}{Z_u}\bigg)-\frac{3\lambda_{12}q_1^2\,p\cdot q_2}{Z_u}\bigg]\notag\\
 &+\frac{6\lambda_{12}q_1^2}{Z_u}\Big(m^2\big(s-q_1^2+q_2^2\big)-P_{21}\Big)-\frac{2q_1^2\lambda_{12}P_{12}}{Z_u}\bigg(1-\frac{4m^2q_1^2}{Z_u}\bigg)\notag\\
 &+\frac{16q_1^2}{Z_u}\bigg(1+\frac{2m^2q_1^2}{Z_u}\bigg)p\cdot q_2\big(s+q_1^2-q_2^2\big)\big(q_1^2\,p\cdot q_2+2P_{12}\big)\Bigg\}\notag\\
 &-\frac{8\lambda_{12}}{45q_1^2}\Big(4m^2q_1^2s+2(p\cdot q_1)^2\big(s+q_1^2-q_2^2\big)-p\cdot q_1\,q_1^2\big(s-q_1^2+q_2^2\big)\Big),\notag\\
 T_{14,s}&=-\frac{128q_2^2P_{12}(P_{12}+P_{21})}{15\lambda_{12}}\Bigg\{2\big(q_1^2+q_2^2\big)-\frac{m^2}{Z_s}\Big(\lambda_{12}-4s\big(q_1^2+q_2^2\big)\Big)\notag\\
 &+\frac{1}{Z_s}\bigg(1+\frac{2m^2 s}{Z_s}\bigg)\bigg[m^2\Big(s^2+\big(q_1^2-q_2^2\big)^2\Big)-4s\,p\cdot q_1\,p\cdot q_2-2\big(q_1^2-q_2^2\big)\Big((p\cdot q_1)^2-(p\cdot q_2)^2\Big)\bigg]\Bigg\},\notag\\
 T_{14,u}&=\frac{8P_{12}\big(4P_{21}(s+q_1^2-q_2^2)+q_2^2\lambda_{12}\big)}{15\lambda_{12}}
 \Bigg\{4q_2^2+\frac{m^2}{Z_u}\big(8q_1^2q_2^2-\lambda_{12}\big)\notag\\
 &+\frac{4}{Z_u}\bigg(1+\frac{2m^2q_1^2}{Z_u}\bigg)\Big[m^2(q_1\cdot q_2)^2-P_{12}\, p\cdot q_2-p\cdot q_1\,p\cdot q_2\, q_1\cdot q_2\Big]\Bigg\},
\end{align}
with the abbreviations
\begin{align}
 P_{12}&=p\cdot q_1\,q_1\cdot q_2-p\cdot q_2\, q_1^2,\qquad P_{21}=p\cdot q_1\,q_2^2-p\cdot q_2\, q_1\cdot q_2,\notag\\
 Z_s&=\big(p\cdot q_1+p\cdot q_2\big)^2-m^2 s,\qquad Z_u=\big(p\cdot q_1\big)^2-m^2 q_1^2.
\end{align}

\end{document}